\documentclass[12pt]{article}
\usepackage[utf8]{inputenc}
\usepackage[T1]{fontenc}
\usepackage{palatino}
\usepackage[english]{babel}
\usepackage[colorlinks=true,urlcolor=black,citecolor=black,linkcolor=black,bookmarks=true, pdfborder={0 0 0}]{hyperref}
\usepackage{graphicx}
\usepackage{microtype}
\usepackage{amsmath}
\usepackage{mathtools}
\usepackage[round]{natbib}
\usepackage[nottoc]{tocbibind}
\usepackage{threeparttable}
\usepackage{longtable}
\usepackage{acronym}
\usepackage{tabularx} 
\usepackage{booktabs}
\usepackage{multirow}
\usepackage[center]{caption}
\usepackage[center]{subfigure}
\usepackage[section]{placeins} 
\usepackage[margin=1in]{geometry}
\usepackage{rotating}
\usepackage{lscape}
\usepackage{adjustbox}
\usepackage{chngcntr}
\usepackage{authblk}
\usepackage{setspace}
\usepackage{fancyhdr}
\title{Can the creation of separate bidding zones within countries create imbalances in PV uptake? \\ Evidence from Sweden.}
\author{Johanna Fink\footnote{Department of Economic History, Lund University, Sweden. johanna.fink@ekh.lu.se}}
\date{\today}

\begin{document}

\maketitle

\begin{abstract}
This paper estimates how electricity price divergence within Sweden has affected incentives to invest in photovoltaic (PV) generation between 2016 and 2022 based on a synthetic control approach. Sweden is chosen as the research subject since it is together with Italy the only EU country with multiple bidding zones and is facing dramatic divergence in electricity prices between low-tariff bidding zones in Northern and high-tariff bidding zones in Southern Sweden since 2020. The results indicate that PV uptake in municipalities located north of the bidding zone border is reduced by 40.9-48\% compared to their Southern counterparts. Based on these results, the creation of separate bidding zones within countries poses a threat to the expansion of PV generation and other renewables since it disincentivizes investment in areas with low electricity prices.
\end{abstract}



\newpage
\linespread{1.2}
\tableofcontents
\doublespacing

\newpage
\section{Introduction}

With the ratification of the \textit{European Green Deal} in 2021, the European Union (EU) manifested its goal to become climate neutral by 2050 \citep{eceee2013}. Even prior to its ratification the European Green Deal was central to the first \textit{State of the Union} speech Ursula von der Leyen as the president of the European commission in 2020. In her speech von der Leyen highlighted fossil-free steel production in Northern Sweden as the center of the sustainable transition of energy intensive industries in Europe \citep{eu2020}. But what makes Northern Sweden so particularly attractive to energy-intensive green industries?

Northern Sweden offers companies access to vast natural resources, such as copper, iron, and rare earth, as well as collaboration possibilities with high level research facilities \citep{smartcity2023}. Nonetheless, \cite{johnson2021} considers the massive surplus of cheap renewable electricity generated from hydro and wind power to be the main pull factor. Industrial consumers in Northern Sweden faced the lowest electric tariff rates in Europe, amounting on some days to merely 0.13 SEK or 1 cent per kWh, a twelfth of the European average \citep{thunborg2021,stromreport2021}. Consequently, Northern Sweden is expected to attract massive investments of approximately 1,000 billion SEK or 120 billion USD over the next decade.

Nonetheless, neither this investment boom nor the low electric tariff rate extend to the entire country. In fact Southern Sweden has faced electricity shortage over the last years, as a consequence of insufficient grid capacity and the shut down of several nuclear power plants \citep{armelius2022elbrist}. While electric tariff rates remained low in Northern Sweden, electricity shortage resulted in soaring electricity prices in the Southern part of the country. Electricity price divergence between Northern and Southern Sweden peaked in August 2022, amounting to a factor of twelve \citep{vattenfall2022}.

This divergence of electricity prices within a country surely appears strange to most inhabitants of Central and Southern Europe where electricity prices are identical throughout the country \citep{entsoe2023}. What makes this divergence possible in Sweden is the division of the country into separate bidding zones (BZ) in 2011. A BZ is defined as \textit{"the largest geographical area within which market participants are able to exchange energy without capacity allocation."} \citep[p. 1]{ofgem2014bidding}. The price within a BZ is determined by the balance of electricity demand and supply, which can result in significant variations in electricity prices across BZs.

The EU considers BZ creation within countries as crucial regarding its efforts to reduce its carbon emissions by promoting the renewable electricity generation \citep{entsoe2023}, from solar, wind, and hydro sources which "have the potential to provide energy services with zero or almost zero emissions of both air pollutants and greenhouse gases.” \citep[p. 1514]{panwar2011role} BZ creation can promote an expansion of renewable in high-tariff BZ with a production deficit since financial factors are generally identified as the main driver of investments in renewable electricity generation \citep{borenstein2017private,palm2020early,gautier2020pv}. 

Moreover, BZ creation is associated with improved grid-stability and a decreased risk of line-overloads \citep{janda2017influence,bergh2016impact}. The latter issue is becoming increasingly important due to seasonal and weather related fluctuation of solar and wind power expansion and has already affected the Central European BZ configuration. Expansion of solar and wind power generation has resulted in the split up of the German-Austrian-Luxembourgian BZ and raised demands by Germany's Eastern European neighbors and North German states to subdivide Germany into multiple BZ \citep{Kurmayer2023,hurta2022impact}. The EU is currently planing to split not only Germany but even France and the Netherlands into sub-national BZ \citep{entsoe2023}. Nevertheless, it remains unclear how this would affect prospects to expand renewable electricity generation in the affected countries, since investment would only be incentiviced in high-tariff BZ.  

This paper aims to estimate the effect of price divergence within countries on PV uptake across BZ in Sweden. The Northern BZ border whose investments in the renewable energy technology is negatively affected by electricity price divergence is at the center of the analysis. To estimate the effect of electricity price divergence on PV uptake, a synthetic control approach is applied, which creates a synthetic comparison unit based on a weighted set of control units. This method is entirely data driven and particularly suited to estimate the impact of policy reforms on macroeconomic units when the pool of potential controls units is limited \citep{abadie2003economic,abadie2021using}.

Sweden is selected as a case study since it is one of only three European countries that is divided into multiple BZ and experienced a dramatic divergence in electricity prices since 2020 between the Northern and Southern BZ \citep{vattenfall2022,armelius2022elbrist}. While Norway experienced a similar shock in electricity prices and consists of five BZ, it is not suited as its solar PV market only emerged since 2020 when the price divergence occurred \citep{owd2023solarshare,owd2023solar}. The choice of studying PV uptake rather than wind power expansion is motivated by the limited time horizon since the electricity price divergence. The construction of wind power plants takes on average 9 years, compared to 4 to 6 month for solar modules in Sweden \citep{blomwestergren2022brist,vind2023tillstand}.

This paper contributes to academic literature by providing important insights on the effect of separate BZ on PV uptake for countries like Norway and Italy that are already divided into BZs but even to other European countries where separate BZs are being established in the near future \citep{entsoe2023}. Moreover, it provides a guideline for the application of the synthetic control method and the selection of border regions for future studies on electricity price divergence.

The remainder of the paper is structured as follows. The second section summarizes existing literature on solar generation, determinants of solar PV uptake, and BZ. The third section provides background information on solar generation in Sweden. In the fourth and fifth section, data sources, alternative border region definitions and the synthetic control method are introduced. The sixth section presents the estimation results. In the seventh section implications of the studies findings and its limitations are discussed. The final section concludes.

\newpage
\section{Review of existing literature}

This section introduces some major distinctions between different modes of solar generation and provides a brief literature review on BZ related research as well as on previous findings on determinants of PV uptake.

\subsection{Bidding zones}

Biddings zones (BZ) are frequently referred to as electric tariff areas since they refer to the largest geographical area with the same electric tariff rate. A BZ is usually defined on the national level but it can even comprise several countries or only part of a country \citep{bergh2016impact,ofgem2014bidding}. 

\begin{figure}[htb!]
    \centering
    \caption{BZ in Europe}
    \includegraphics[scale=0.7]{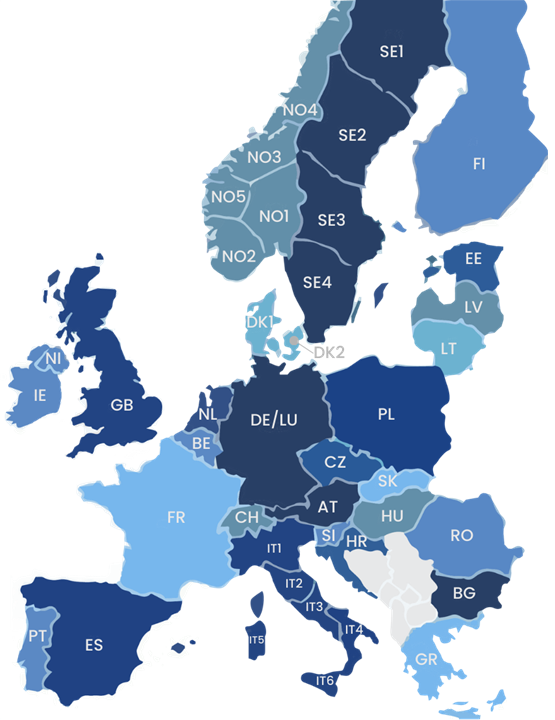}
    \label{fig:european_bz} \\
    Source: \cite{oliveira2023}
\end{figure}

The European BZ configuration is presented in figure \ref{fig:european_bz}. Most countries such as France, Spain and the Netherlands consist of a single BZ. Sweden is along with Norway, and Italy, one of merely three European countries with multiple BZ. BZ can stretch over multiple countries such as in the case of Irish BZ, which comprises British and Irish territory, or the German-Luxembourgian BZ. The latter is a special case since it comprised even Austria until the split of the tri-national BZ in 2018 \citep{hurta2022impact}.

Much of the existing research on BZ is focused around topics like network stability and optimal BZ configuration. \cite{bergh2016impact} quantify the impact of the number of BZs on electricity market outcomes of the Central European electricity system. Their findings suggest that a rising number of BZ is positively associated with network characteristics and a reduced risk of line overloads.

\cite{breuer2014optimized} focus their research on the most efficient BZ.
The authors fund that electric networks operate most efficiently if BZ borders reflect transmission bottlenecks. Since networks evolve over time \citeauthor{breuer2014optimized} suggest that BZ borders should be regularly readjusted to account for changes in network structure and the geographical allocation of electricity supply and demand. Nonetheless, this would have negative implications for planning security of market participants and harm investments into renewable generation capacity. Especially if there exist substantial price differences across BZ, changing the BZ borders would alter amortization periods. 

Additionally, there exists a vivid discussion on the impact of BZ creation on the possible reallocation of energy intensive industries \citep{tagesschau2022,entsoe2023}. This is an especially fierce topic in Germany, where the surplus generating Northern states demand the creation of separate BZ. This would allow locals to benefit from previous investments into renewable generation capacity in form of reduced tariff rates and attract energy intensive industries to the otherwise agriculturally dominated North of Germany. Southern German states oppose this idea since they fear a reallocation of energy intensive industries to electricity abundant areas.

The effect of price divergence across BZ on investments in renewable energy technology has so far only been addressed in two studies. \cite{hurta2022impact} estimate how the split of the German-Austrian-Luxembourgian BZ in 2018 has affected investments in battery storage systems, a popular tool to balance electricity supply variations from wind and solar generation. The authors use a flow model and data from 2 years before and after the BZ split for their estimation. \citeauthor{hurta2022impact}'s finding reveal that the higher electricity price volatility in Germany raises incentives for investments in battery storage systems the German-Luxembourgian relative to the Austrian BZ, since Austria relies mostly on hydropower and biofuels which can be adjusted to electricity consumption. 

A limitation of \citeauthor{hurta2022impact}'s (\citeyear{hurta2022impact}) study is that actors that are located in the different BZ are even located in different countries. Therefore they do not face the same incentives schemes nor national policies, affecting their investment decisions. To circumvent this issue price divergence across BZ within a single country should be exploited. There currently exists merely one example for this type of study by \cite{mauritzen2023great}, who investigates the impact of electricity price divergence between Northern and Southern Norwegian BZ on pure-electric vehicle uptake. Similar to the Swedish case, Southern Norway has experienced electricity shortage and therefore a price increase compared to the North of the country. To further reduce the issue of underlying differences in investment propensity across BZ, \citeauthor{mauritzen2023great} merely considers municipalities close to the border between the high- and low-tariff BZs. It should be noted, that the paper lacks a clear motivation which municipalities should be part included in the border region.

Based on a simple difference and difference-in-difference approach, \citeauthor{mauritzen2023great} estimates that Southern Norwegian consumers were 2-5\% less likely to invest in electric vehicles compared to their Northern Norwegian counterparts. This relatively modest impact can be accounted to the long-term nature of car purchases, the expectation that electric tariff differences will not persist in the long run, and the Norwegian ban of non-electric vehicle sales by 2025.

\subsection{Photovoltaic generation}

There are two main approaches to use sun light for energy generation, solar thermal and photovoltaic (PV) generation. This paper focuses on solar PV which "generate[s] electricity directly from light without emissions, noise, or vibration" \citep[p. 1516]{panwar2011role}. PV generation requires sunlight needs to shine on solar modules made of semiconductor materials \citep{satpute2020mercurochrome,lee2017review}. Silicon-based crystalline solar modules are the by far the most popular, accounting for 95\% of the world market \citep{andersson2021photovoltaics}. Alternatives like thin-film applications are either made from amorphous silicon, copper indium gallium selenide, or cadmium telluride \citep{lee2017review}. Despite lower material requirements and superior performance in highly insolated areas thin-film applications fail so far to compete with crystalline solar panels.

Apart from different types of solar modules, PV application can be divided into on- and off-grid applications \citep{andersson2021photovoltaics,silva2016electricity}. The latter are of major importance for electrification of islands, boats and remote areas. In developed countries such as Sweden, off-grid applications were especially popular during the early stages of PV adoption to electrify remote summer houses and boats. With increasing adoption of PV installations this generation mode lost popularity accounting for merely 0.4\% of installed PV capacity in 2020 \citep{energimyndigheten2020}. 

On-grid applications can be further sub-divided into centralized and distributed generation. \citep{andersson2021photovoltaics}. On-grid centralized PV generation is the most popular solar application, accounting for 55\% of generation capacity worldwide in 2021 \citep{iea2022ppsp}. Nonetheless, this generation mode is less popular in Europe especially in Sweden, were the share of distributed generation amounts to 94\%. The main advantages of roof-mounted distributed PV applications over centralized plants are reduced transmission losses and improved network stability, since electricity is produced where it is consumed \citep{tamimi2013system}. Moreover, roof-mounted solar installation cause less social tension. Rising electricity prices have incentivized farmers to convert productive agricultural land into centralized solar farms creating a conflict between energy and food production and evoking resistance against centralized solar generation in Europe and the US \citep{Katanich2023,zinke2021,santos2019challenges}. 

A main concern regarding increased PV expansion are fluctuations in PV generation capacity, which is affected meteorological and seasonal fluctuations as well as topological aspects \citep{castillo2016assessment}. PV potential is greatest in the flat land areas close to the equator where seasonal changes in solar irradiation are least prominent. PV capacity is limited in mountainous areas where hills block solar irradiation in valleys most of the day. Further, PV capacity is reduced in areas with high occurrence of clouds and precipitation rates.

To balance electricity generated from solar sources over time, PV capacity need to be backed up with other forms of electric generation or coupled with storage facilities, typically in the form of batteries \citep{schmidt2016optimal}. The latter is of special importance in off-grid systems where solar energy serves as the sole energy source. However, integrated system including battery storage units are considerably more expensive limiting the ability of low-income households to afford solar installations. Additionally, batteries production is energy intensive based and disposal of toxic substances from battery production harms the environment \citep{mcmanus2012environmental}.

\subsection{Determinants of PV uptake}

The selection of control variables for this study, requires and understanding of the main determinants of PV uptake. Determinants of PV uptake can be broadly categorized as geographic and socio-economic determinants which are discussed in separate subsections. 

This review summarizes empirical findings from the US \citep{borenstein2017private,graziano2015spatial}, Japan \citep{zhang2011impact}, Germany \citep{winter2019german,baginski2019coherent}, Belgium \citep{gautier2020pv}, Malta \citep{briguglio2017households}, Brazil \citep{assunccao2017developing} and the UK \citep{balta2015regional}.

\subsubsection{Socio-economic factors}

Findings of \cite{balta2015regional}, \cite{borenstein2017private}, \cite{graziano2015spatial}, \cite{gautier2020pv}, and \cite{assunccao2017developing} suggest that economic and financial incentives in form of electricity tariffs, disposable income, and unemployment are the main determinants of PV investments. \cite{gautier2020pv} estimate that a marginal increase of the electric tariff rate by 0.01€ per kWh increase PV uptake by 8\%. Furthermore, unemployment is associated with economic instability hampering the general propensity to invest. \cite{briguglio2017households} estimate that a 1\% increase in the unemployment rate reduces the share of households with PV systems by 0.2\%. Moreover, the financial situation of households has important implications for PV uptake. \cite{borenstein2017private} and \cite{winter2019german} detect a strong positive correlation between disposable household income and investment in solar generation since poor households frequently lack the financial means to cover initial investment costs. Since the initial investment costs are a major impediment to PV uptake, several governments offer assistance for households to cover initial investment costs \citep{pvmagazin2022}. This kind of subsidies exist among others in Austria and Cyprus. 

Moreover, most European governments have implemented incentive schemes to promote renewable generation \citep{pvmagazin2022,bergek2010tradable}. \textit{Feed-in-tariffs} are the most popular way to promote investment in renewable generation capacity. Under this scheme, renewable electricity producers can consume their own electricity and sell their excess production on the electricity market at a fixed rate, which compensates them for generation costs but also provides them with an economic return. 

The most popular incentive system in Scandinavia are \textit{tradable green certificates} (TGC) \citep{bergek2010tradable,iea2019,pvmagazin2022}. As in countries with a feed-in tariff, renewable electricity producers can under a TGC scheme consume their self-generated electricity for free. Nonetheless, they do not receive a fixed amount of money per kWh inserted into the electric grid. Instead, investment costs are compensated through a the electricity market price plus a TGC, which is determined on a separate market. Governments create demand for TGC by setting requirements for providers to source a specified amount of electricity from renewable sources. To ensure an expansion of renewable generation, the minimum required share is regularly increased, raising the demand and thereby also the value of TGCs.

According to \cite{bergek2010tradable} and \cite{fridolfsson2013reexamination}, TGC are an effective tool to raise the share of renewable generation for little tax funds. Nevertheless, 
TGC are problematic in regard to their effect on equity as well as technological development. Equity concerns arise since all producers of renewable electricity benefit from TGC including incumbent actors which do not face additional investment costs, providing them with access profit. Moreover, relying on a TGC therefore impedes the incorporation of novel electricity sources into a countries energy mix. Since the price of TGC is independent from the energy source, investors generate the greatest returns by investing in the technology with the lowest marginal costs. Relying on a TGC therefore impedes the incorporation of novel electricity sources into a countries energy mix.

Further socio-economic determinants of PV uptake include average age, environmental behavior and education levels which are however not considered as control variables in this study. Education is generally positively associated with PV uptake \citep{briguglio2017households,balta2015regional}. Nevertheless, since there also exists evidence for a correlation between household income and education level \citep{fischer2021europe}, the inclusion of both variables might bias estimation results. Empirical findings regarding the impact on age and environmental behavior point into different directions or lack statistical significance. \cite{gautier2020pv} estimate higher PV uptake among older households in Wallonia, while it is most popular among young Maltese households \citep{briguglio2017households}. 

Environmental behavior reflected has been tested in several studies \citep{briguglio2017households,gautier2020pv,baginski2019coherent}, but merely \cite{zhang2011impact} find a significant positive effect on PV uptake. This can be explained by the observation period in the study by \cite{zhang2011impact} which dates back to the early 2000s, when solar generation was generally not economically viable \citep{irena2021}. According to \cite{palm2020early}, the importance of environmental behavior diminishes as PV investments become financially attractive.

\subsubsection{Geographic factors}

The first of the geographic determinants is solar irradiation, which is affected by the distance to the equator, daytime, seasons and local weather conditions and topography \citep{yang2020potential,castillo2016assessment}. While solar radiation increases the productivity of PV modules, empirical results regarding the relationship between solar irradiation and PV uptake are very mixed. \cite{winter2019german} and \cite{baginski2019coherent} detect a positive relationship between insolation rates and PV uptake in Germany. According to \cite{baginski2019coherent} a 10\% increase in irradiation translates to a 2.3\% increase in installed potential. Nonetheless, \cite{zhang2011impact} find no significant effect of irradiation in PV uptake in Japan, and in Brazil the relationship is even reversed \citep{assunccao2017developing}.

Another factor that should be considered is the housing structure. \cite{winter2019german} and \cite{briguglio2017households} identify roof space per capita as a key determinant of PV investment since PV installations are mainly worthwhile if they can cover the electricity consumption of the houses residents. This corresponds to the findings of \cite{graziano2015spatial} whose findings suggest that PV uptake is highest in small and medium sized communities, which are especially in the US dominated by single family homes.

Access to complementary infrastructure and knowledge on alternative energy sources can be captured as neighborhood effects. According to by \cite{balta2015regional} and \cite{graziano2015spatial}, neighborhood effects raise PV uptake in the short-run as greater awareness of the technology and the presence of solar installers increases households likeliness to invest in solar installations. However, this effect evaporates over time as market saturation is approached. 

\newpage
\section{The Swedish case}

Diffusion of solar generation is considered by several Swedish authorities as crucial for achieving Sweden's climate targets and promote a transition towards entirely renewable electricity generation \citep{regeringskansliet2015,miljodepartementet2019,miljodepartementet2020}. This section provides background information on the history of solar generation and research in Sweden, as well as the establishment of separate BZ in the country.

\subsection{The Swedish PV market}

Off-grid solar generation emerged in Sweden already during the 1990s with the aim to electrify boats and remote cottages \citep{andersson2021photovoltaics}. The government has supported this mode of solar generation actively with the establishment of the national development project \textit{SolEl} which specifically targeted off-grid applications in 1995. From 2000 until SolEl's discontinuation in 2018 it promoted even on-grid applications \citep{energimyndigheten2013solel}. 

\begin{figure}
    \centering
    \caption{Grid connected PV installations by  plant size, 2016-2022}
    \includegraphics[width=.43\linewidth]{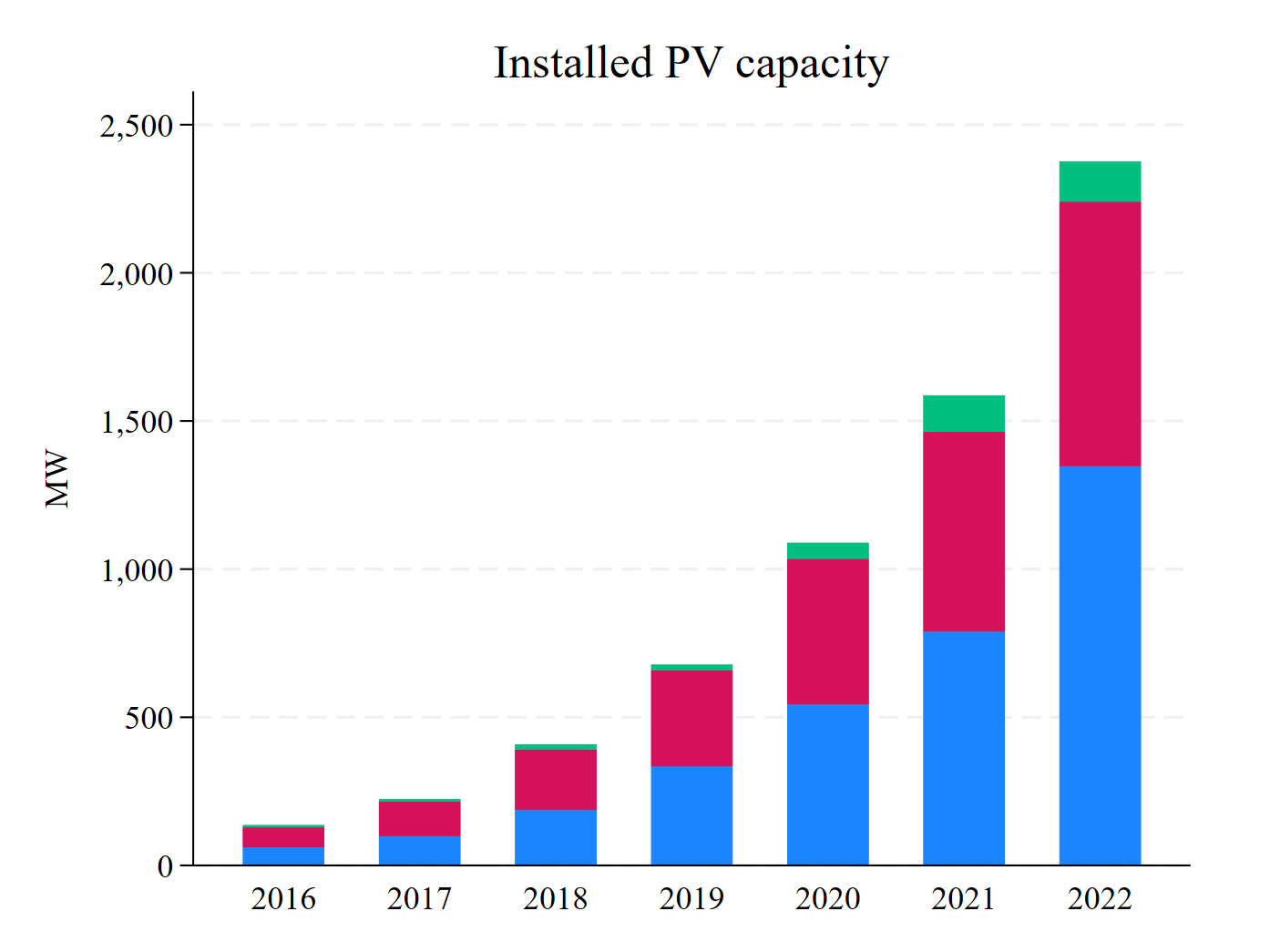}
    \label{fig:units}\includegraphics[width=.55\linewidth]{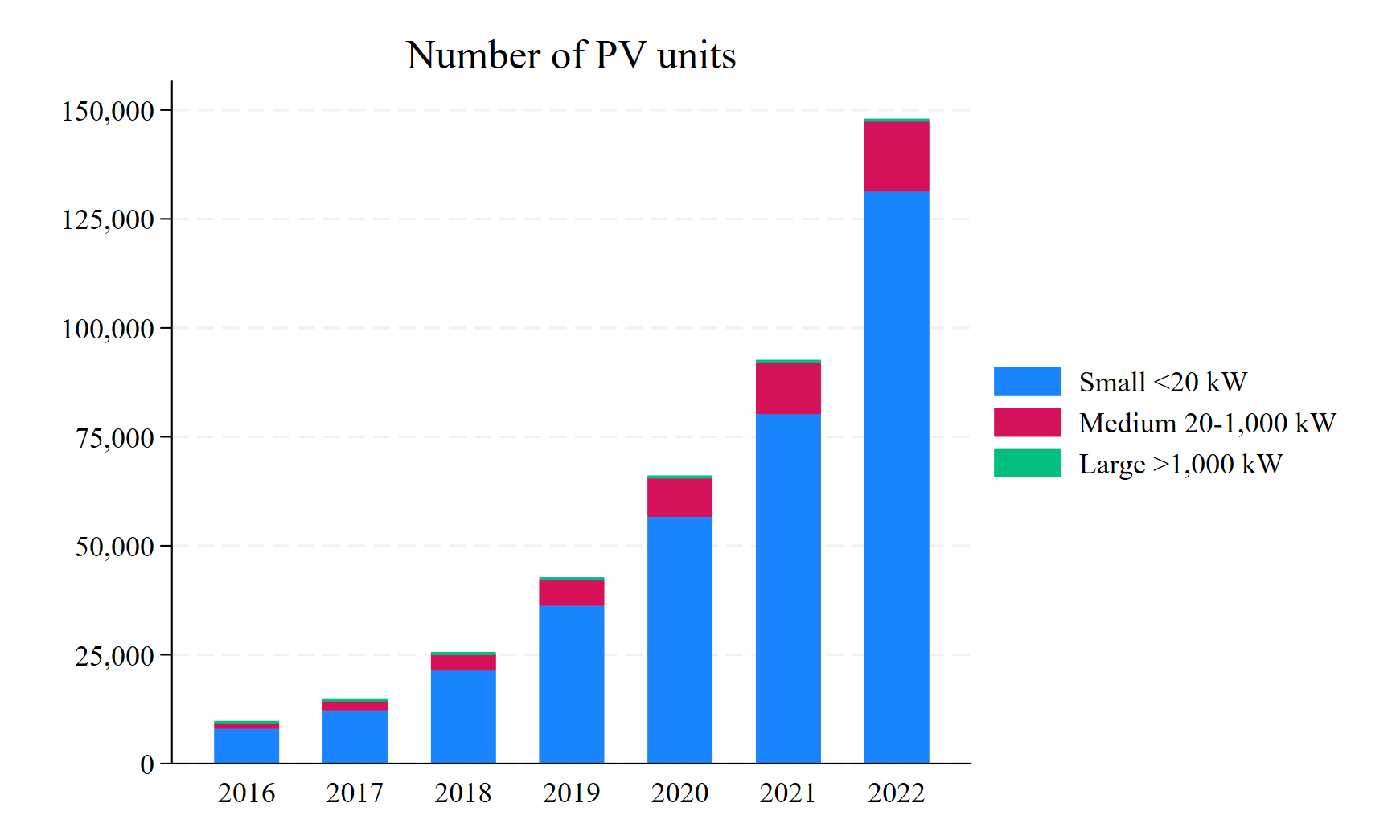} \\
    Source: \cite{energimyndigheten2022}
\end{figure}

Despite the emergence of solar generation in the mid-1990s, it should take two more decades until diffusion of solar generation occurred on the large scale \citep{energimyndigheten2022,andersson2021photovoltaics}. As illustrated in figure \ref{fig:units}, Sweden experienced exponential growth in PV installations since 2016 and does not display any signs of saturation. Between 2016 and 2022 total installed capacity increased by more than a factor of 17 reaching 2.38 GW by the end of 2022. This is far below the by \cite{yang2020potential} estimated 65-84 GW of potential capacity of roof-mounted solar installations in Sweden.

\begin{figure}
    \centering
    \caption{Share of solar PV in total electricity generation, 2000-2022 (\%)}
    \includegraphics[scale=0.45]{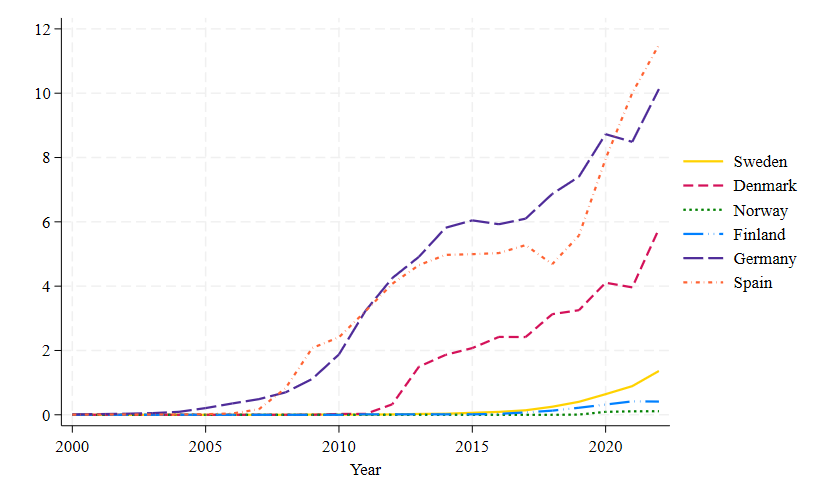}
    \label{fig:solar_share} \\
    Source: \cite{owd2023solarshare}
\end{figure}

Nonetheless, the share of PV generation in Sweden remains low in an international comparison. The share of solar in total electricity generation between 2000 and 2022 is illustrated in Figure \ref{fig:solar_share}. While PV accounted for 1.36\% of total electricity generation in Sweden, the share in Denmark was close to 6\% and exceeded 10\% in Germany and Spain. Only the neighboring Scandinavian countries Norway and Finland displayed significantly lower levels of PV generation amounting to merely 0.1\% and 0.4\%, respectively.

The comparably late investment in Sweden can be accounted to the countries geography but also financial incentives for investment in PV modules. First, Sweden is located between the 55\textdegree and 70\textdegree North, resulting in lower average radiation and higher seasonal fluctuations compared to countries with higher PV uptake \citep{castillo2016assessment,smhi2022}. 

\begin{figure}
    \centering
    \caption{Levelized renewable energy generation costs, 2010-2020 (2020 USD/kWh)}
    \includegraphics[scale=0.2]{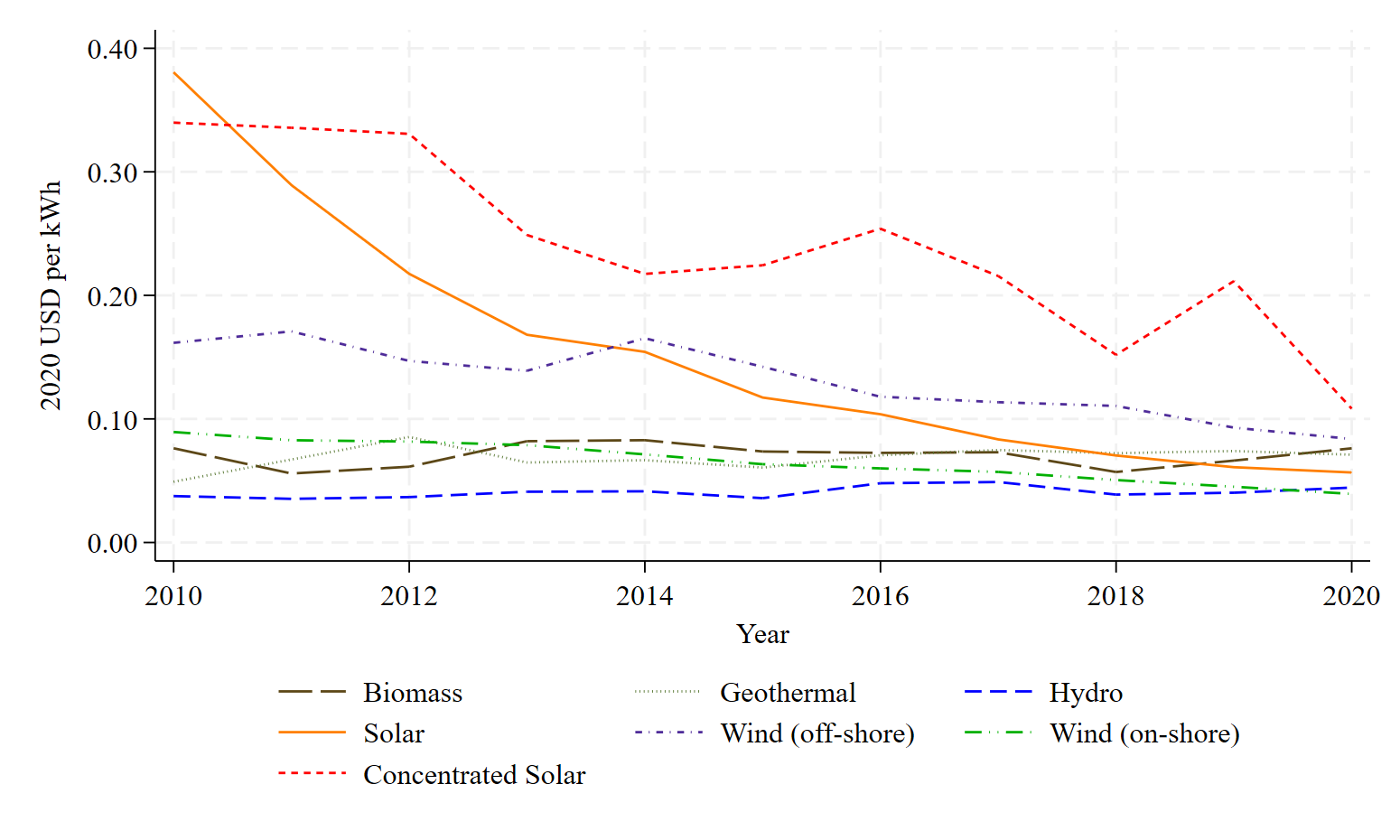} \\ \vspace{0.2cm}
    \footnotesize Source: \cite{irena2021}
    \label{fig:generation costs}
\end{figure}

Second, Sweden utilizes a TGC to promote renewable generation which primarily promotes the diffusion of mature technologies with low marginal costs (see section 2.3.1). As illustrated in figure \ref{fig:generation costs} levelized generation costs of solar generation exceeded those of the cheapest renewable source hydropower by a factor of 10 in 2010. Since then the price of solar modules has fallen dramatically dropping to 0.06 USD per kWh in 2020, which exceeds the marginal costs of the cheapest renewable sources. Consequently PV diffusion was impeded until the mid-2010s, when PV installations became cost competitive with alternative energy carriers \citep{bergek2010tradable,iea2019,pvmagazin2022}. PV generation is not the only energy technology impeded by the TGC. Wind power is the second most important renewable energy source in Sweden accounting for 17\% of total electricity generation in Sweden but its diffusion was delayed until 2010 when it became cost competitive with alternative technologies \citep{iea2022,bergek2010wind}. Consequently, the TGC is likely to impede the diffusion of energy carriers developed in the future.

Third, while the first government subsidy for PV installations was very generous covering 70\% of initial investment costs, wide-spread was impeded since it merely targeted public actors until 2009 \citep{andersson2021photovoltaics}. In 2009, this scheme was extended to all on-grid distributed generation, promoting PV investment throughout the population. Together with falling marginal costs PV capacity increased by 53\% annually but contributed still only for 0.1\% of total generation capacity by 2015. With falling marginal costs and rising PV uptake over recent years, investment subsidies have undergone additional reforms under recent years and became overall less generous. Since 2019, it is limited to 20\% of installation costs or 1.2 million SEK per plant \citep{swedishgov2020integratedplan}. Additionally, the government covered 60\% of initial investment costs for storage units. In December 2020 subsidies for initial investment costs were discontinued and fully replaced with tax exemptions, amounting to 20\% and 50\% for PV modules and storage systems, respectively \citep{energimyndigheten2023solelportalen}.

Fourth, Sweden lacks linkages between PV research and production. Since the emergence of solar research in the 1980s, Swedish PV research in focused on thin-film PV modules panels, an alternative to the dominant crystalline solar panels \citep{andersson2021photovoltaics}. While the first PV company Solibro specialized on this technology, other PV companies that started production in the early 2000s focused on the more popular crystalline modules instead \citep{energimyndigheten2020}. The situation deteriorated when competition from China resulted in the collapse of the European PV industry in the late 2010s. Solar cell production literally disappeared in Sweden by 2016. Since 2019, two companies started to produce thin-film PV modules. However, their combined production of 1.2MW is by far insufficient to cover the demand for new solar installations in Sweden which exceeded 400MW in the same year \citep{energimyndigheten2020,energimyndigheten2022}.

\subsection{Electricity price divergence}

\begin{figure}[htb!]
    \centering
    \caption{Electric tariff areas}
    \includegraphics[scale=0.6]{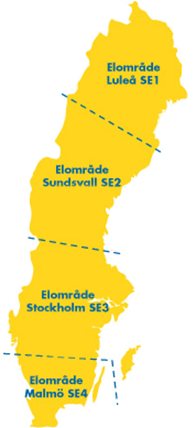} \hspace{2cm}
    \includegraphics[scale=0.3]{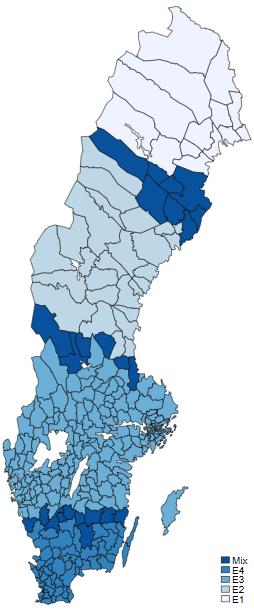}
    \label{fig:elomrade} \\
    Source: \cite{she2022elomraden} and \cite{svenskakraftnat2022}
\end{figure}

Sweden was divided into four BZ in 2011, following a directive from the EU with the purpose to prevent sudden export stops of Swedish electricity to other EU members \cite{ei2022}. The Swedish BZ configuration is illustrated in figure \ref{fig:elomrade}. BZ do not necessarily follow municipality borders. In fact 25 municipalities marked in dark blue are located in multiple BZ allowing for tariff divergence within a single municipality. A complete list over the Swedish municipalities and the tariff area they belong to can be found in table \ref{tab:municipality_list} in the appendix.

\begin{figure}[htb!]
    \centering
    \caption{Floating electricity tariff by BZ, 2015-2022 (öre/kWh)}
    \includegraphics[scale=0.2]{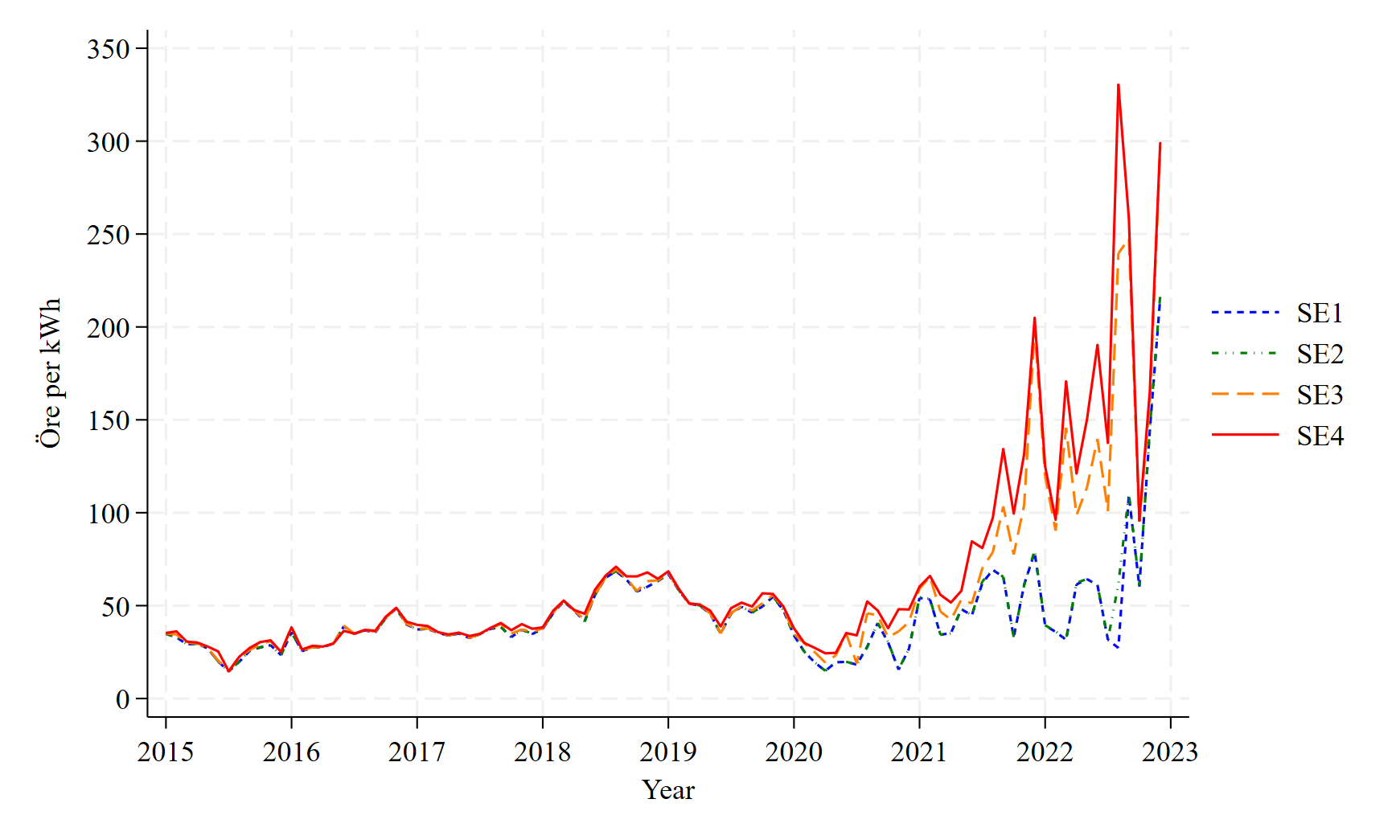}
    \label{fig:tariff} \\
    Source: \cite{vattenfall2022} 
\end{figure}

By creating different tariff areas policy makers opened even the doors for differentiated electricity prices across tariff areas. Electricity prices between 2015 and 2022 are illustrated in figure \ref{fig:tariff}. While electricity prices fluctuated over time, divergence across BZ emerged only in 2020 when electricity shortages in Southern Sweden drove up electricity prices in BZ SE3 and SE4. Creating the conditions for a natural experiment on solar PV uptake along the border of the high- and low-tariff BZ SE3 and SE2. According to \cite{vattenfall2022}, electricity prices differences reached a record of a factor of 12 in August 2022. Since mid-2022 electricity prices started to converge as prices in the Northern BZ SE1 and SE2 increased.

\begin{figure}
    \centering
    \caption{Installed PV capacity per capita by BZ, 2016-2022 (kW)}
    \includegraphics[scale=0.2]{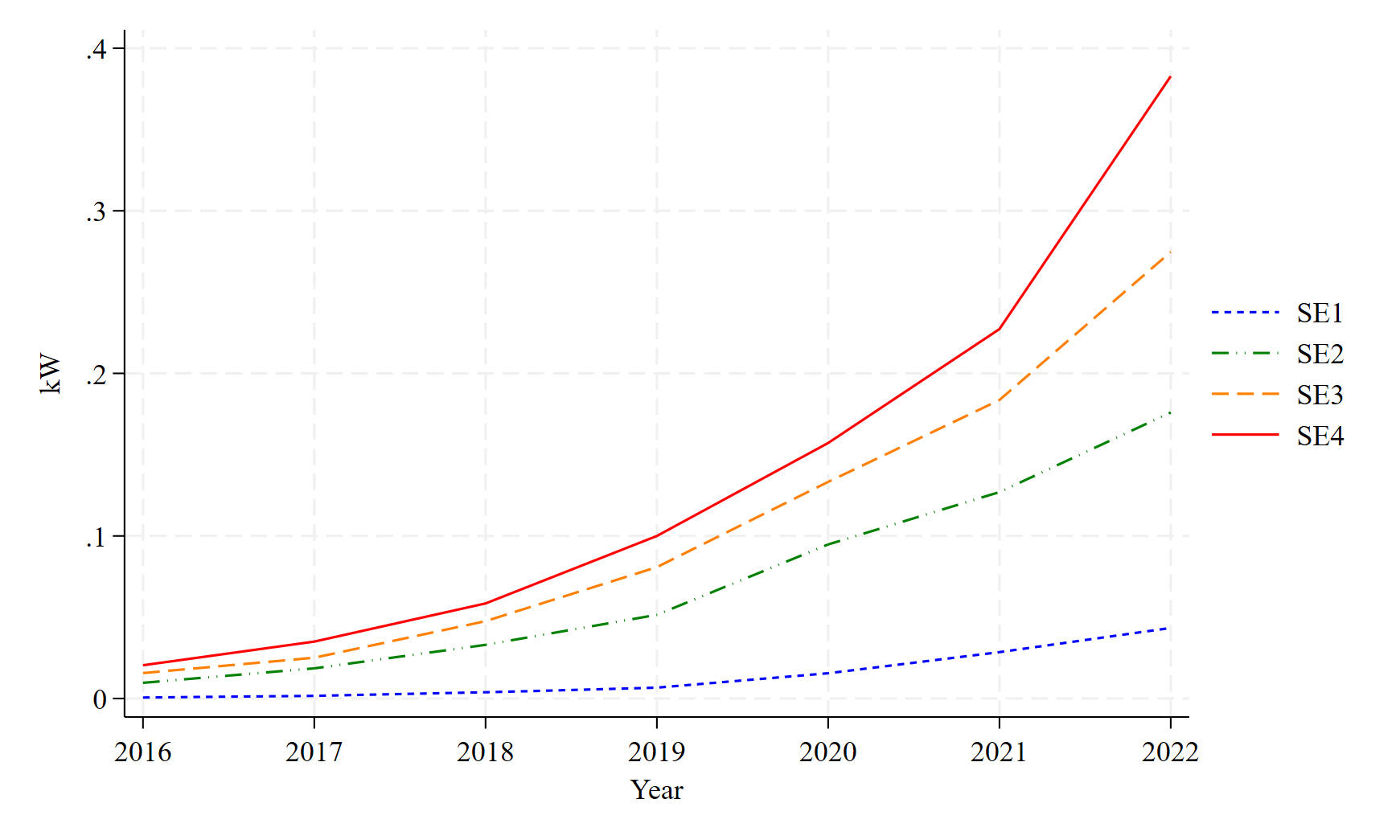} \\ \vspace{0.2cm}
    \footnotesize Source: \cite{energimyndigheten2022}
    \label{fig:capacity_by_bz}
\end{figure}

Installed PV capacity per BZ is illustrated in figure \ref{fig:capacity_by_bz}. PV uptake in Sweden is highest in the Southern most BZ SE4 and lowest in SE1 in the far North. While this corresponds to the higher productivity of PV installations in the South of the country the gap has widened since 2020, when rising electricity prices drove up demand for PV installations in the Southern BZ. The aim of this study is to isolate the effect of electric tariff divergence on PV uptake from other factors impeding PV uptake in Northern Sweden.

\newpage
\section{Data}

To estimate the effect of price divergence across BZ on PV adoption, a balanced panel data set is created containing a total of 2,030 annual observations for all 290 Swedish municipalities between 2016 and 2022.

The choice of the observation period is driven by data availability of the outcome variable installed PV capacity is provided by \cite{energimyndigheten2022} on the municipality level only from 2016 onwards. It should be noted that only small- and medium-sized on-grid installations with a maximum capacity of 1 MW are considered. This is motivated by the fact that these account for 95\% of the solar PV market in Sweden, and that there existed merely 11 PV plants with a capacity above 1 MW which accounted for less than 3\% of solar capacity by the end of 2019 \citep{energimyndigheten2020,energimyndigheten2022}. 

Moreover, installed PV capacity is measured in per capita terms to account for the vast differences in population size across Swedish municipalities which has significant implications for electricity demand as these two factors are positively correlated and demand for PV systems \citep{assunccao2017developing}. In 2021, Swedish population by municipality varied from merely 2,385 residents in Bjurholm (Västerbotten) to approximately 1 million in Stockholm \citep{scb2022}.

Since the main aim of this paper is to estimate how price divergence across BZ affects PV uptake, the treatment status is determined by a dummy variable which takes the value 1 if a municipality is located in a low-tariff BZ SE1 or SE2. The seven municipalities that cannot clearly be assigned to the low- or high-tariff areas are excluded from the sample. The treatment status of all municipalities is documented in table \ref{tab:municipality_list}.

Additionally, solar irradiation, unemployment rates, disposable income, and the housing structure are included as control variables. Irradiation data is sourced from the Strång database which is provided by the Swedish Meteorological and Hydrological Institute \citep{smhi2023strang}. Data from Strång can be extracted for every single geographic location in Scandinavia since 1999. For this paper daily global radiation data is extracted for each communities central settlement (\textit{centralort}) over the period 2010-2015. 

The obtained irradiation data is used to create two variables. Average radiation refers simply to the average daily radiation between 2010-2015. Radiation variation is added to account for seasonal variation in sunlight hours which reduces incentives for the adoption of PV systems in areas with large seasonal variations \citep{castillo2016assessment}. It is calculated by the average radiation per month over the 6-year period and then dividing the maximum by the minimum monthly radiation. The value of both variables is fixed within municipalities over the observation period.

Information on disposable income, population size, and the housing structure of Swedish municipalities is retrieved from the \cite{scb2022}. Annual observations of disposable income are reported in per capita terms for all Swedish municipalities from 2000 onwards.

Housing structure refers to the share of small houses. The \cite{scb2022} distinguishes between small houses (\textit{småhus}), with 1-2 housing units, apartment buildings (\textit{flerbostadshus}) with 3 or more housing units, other houses (\textit{övriga hus}) which do not exclusively serve for housing purpose and special housing (\textit{specialbostäder}) like student housing and retirement homes. According to \cite{ohrlund2020rising} it is important to consider the housing structure when studying solar PV uptake in Sweden. PV diffusion among residents of housing cooperatives and rental buildings is slowed down or prevented since they must align their investment decisions. Moreover, roof area per inhabitant is smaller compared to single family houses reducing the attractiveness of PV investments \citep{winter2019german}.

The unemployment data is merely available at the regional level (\textit{län}). The data measures unemployment as the proportion of people in the labor force aged 15-74 years who are unemployed on the regional level. Important to note is that unemployment data is missing for Gotland after 2019. It is assumed that unemployment stayed constant at the 2019-level of 6\% after this.

To avoid the risk of bad controls observations of the time-variant control variables are assigned to the following year. Hence, the controls for the 2016 observations are measured in 2015. This is motivated by the risk of bad control variables. If control variables are determined simultaneously or after the outcome variable the value of the control variable might be determined by the outcome variable instead \citep[p.47-48]{angrist2009mostly}. 

\begin{table}[htb!]
\centering
\caption{Descriptive statistics}
\begin{tabular}{lcccccc}
\hline
Variable     & Observations   & Unit   & Mean    & Std. dev. & Min     & Max     \\ \hline
Installed capacity     & 2,030 & kW  & 0.105     & 0.114       & 0       & 0.95 \\
per capita     &     &     &     &     &     &  \\
Average radiation  & 2,030 & kWh/m\(^2\) & 1,240 & 96 & 931 & 1,437   \\
Radiation variation  &  2,030   &     &  42.3   &  71.7   &   15.7  &  764.6\\
Disposable income  & 2,030 & SEK    & 210,002 & 25,028    & 165,000 & 463,000 \\
Unemployment & 2,030 & \%     & 7.0     & 1.6       & 4.3     & 11.5    \\
Housing structure & 2,030 & \%     & 61.9    & 0.1       & 1.4     & 90.2    \\
 \hline
\end{tabular}                   
\label{tab:descriptives}
\end{table}

Descriptive statistics are presented in table \ref{tab:descriptives}. There exists considerable variation in outcome and control variables across municipalities. While average installed capacity amounted to 0.1 kW per capita over the entire observation period, the maximum installed capacity of 0.95 kW per capita was reported in Borgholm (Kalmar) while three municipalities in Norbotten had not even a single PV installation by 2020.

Another variable that stands out are radiation levels, which were highest in 
Karlskrona (Blekinge) and lowest in Kiruna (Norbotten), amounting to 1,437 and 931 kWh/m\(^2\), respectively. Kiruna even has the highest radiation variation of 765, due to its location North of the Arctic circle \citep{smhi2022}.

\subsection{The three border regions}

As discussed previously, PV uptake across BZ differs substantially across Sweden. This can partly be accounted to underlying differences in control variables, in particular solar radiation levels which are significantly lower in the low-tariff BZ in Northern Sweden. Simply comparing BZ to one another would therefore most likely not produce any meaningful results.

Instead, the analysis focuses on the border region between the low- and high-tariff BZ SE2 and SE3. A similar approach has been applied by \cite{mauritzen2023great} to study electric vehicle uptake in Norway. Nonetheless, selection criteria for including municipalities in the border region remain somewhat unclear apart from the fact that neither of them contains any major cities.

\begin{figure}[htb!]
    \centering
    \caption{Border regions}
    \includegraphics[scale=0.25]{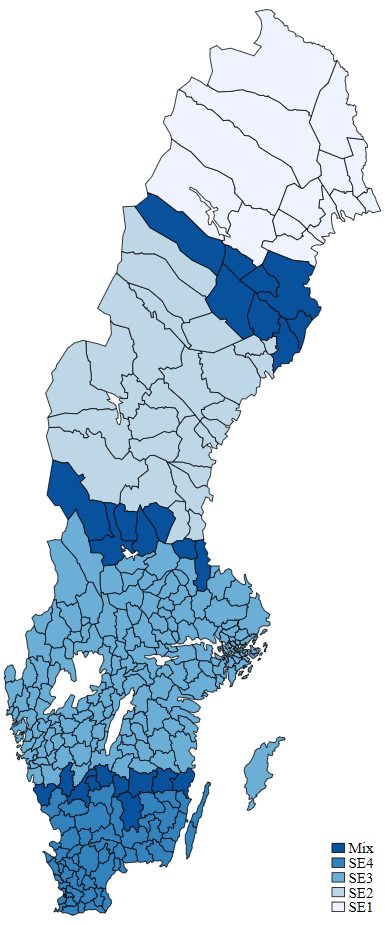} \hspace{2.5cm}
    \includegraphics[scale=0.25]{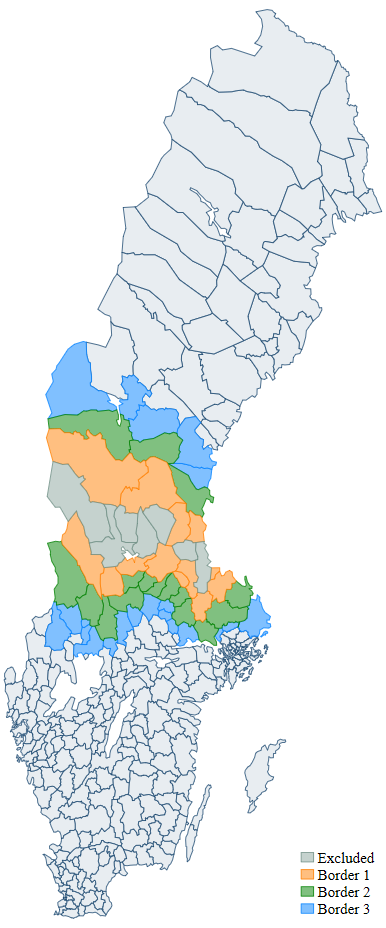}
    \label{fig:border_regions}
\end{figure}

In the following analysis three different definitions of the BZ border are considered, defined by their geographical difference from the border between the low- and high-tariff BZ. The resulting tariff areas are presented in figure \ref{fig:border_regions}. A list over the municipalities included in the different border regions is included in table \ref{tab:municipality_list} in the appendix.

Border 1 refers to municipalities at the BZ border between electricity tariff areas SE2 and SE3 or those municipalities that are part of both areas and therefore excluded from the analysis. Border 2 comprises all municipalities included in border 1 plus municipalities that have a direct border with any of the municipalities in the border 1 group. Similarly, border 3 includes all border 2 municipalities as well as those that have a direct border with them.

The data on outcome and control variables for each border region is calculated by taking the average of outcome and control variables. Since municipalities located in the high- and low-tariff areas are considered separately, values for six different border regions are calculated. The three Northern BZ border regions are of main interest for the following analysis.

\newpage
\section{Synthetic Control Method}

The Synthetic Control Method (SCM) has been developed by \cite{abadie2003economic} as an alternative to time-series analysis, comparative case studies, and difference-in-difference (DiD) approaches which are commonly used to quantify the effects of policy interventions by\citep{card1990impact,richardson2009monetary}. 

To estimate the actual treatment effect the outcomes for the affected unit should be compared in the presence and absence of treatment, which refers here to the location of a municipality in the low-tariff BZ. The problem however is that a municipality cannot simultaneously face high- and low-tariff prices. Hence, it is necessary to model the counterfactual outcome, the one that cannot be observed with help of econometric techniques \citep{billmeier2013assessing,abadie2003economic,abadie2021using}.

The synthetic control group (SCG) is defined as the "weighted average of the units in the donor pool" \citep[p.395]{abadie2021using}. The donor pool consists of non-treated comparison units. The treated unit in this study is the Northern BZ border located in the low-tariff BZ SE2, while the donor pool consists of all municipalities in the high tariff areas SE3 and SE4.

\begin{equation}
    \mathrm{Y^S_{it,T=0}=\sum^{J+1}_{j=2}w_jY_{jt,T=0}}
    \label{eq:scg}
\end{equation}

The composition of the SCG \(Y^S\) of the treated unit \textit{i} is formally presented in equation \ref{eq:scg}. Observations from the donor pool \textit{j} are assigned  weights \textit{w} to minimize the difference between the pre-treatment outcomes of the treated and the SCG, presented in equation \ref{eq:scg}. 

Weights are assigned based on a set of predictor variables, whose selection is similar to the selection of control variables in other econometric models a fundamental component of the estimation process \cite{abadie2021using}. Predictor variables consist of pre-intervention values of the outcome variable combined with variables that potentially affect outcome levels. In this study the predictor variables consist of three pre-treatment values of the outcome variables installed PV capacity per capita in 2016, 2018, and 2020, average radiation, radiation variation, disposable income, the share of small houses, and the unemployment rate.

The SCM rests on two main assumptions. First, according to the non-interference assumption the treatment status of treated unit \textit{i} is independent from donor units \textit{j}. Second weights are restricted to non-negative values and must sum up to one. This ensures a donor pool consisting of relatively few comparison units \citep{abadie2003economic,abadie2021using}. 

The Synthetic control approach has become popular over recent years since it has several advantages over other comparative case study designs. First, SCM are particularly suited when only few control units are available or when outcomes for aggregate units such as municipalities or countries are estimated \citep{abadie2010synthetic,abadie2015comparative,abadie2021using}.\cite{abadie2015comparative} apply SCM to estimate the effect of reunification on West German GDP per capita, while \cite{dube2015pooling} estimate the effect of minimum wage policies across US states. 

Second, SCM offers in difference to comparative case studies and DiD a formal and entirely data driven selection procedure to identify control units. Thereby SCM limits researchers' possibilities to bias estimation results through p-hacking or specific searches \citep{billmeier2013assessing}. Moreover, SCM can even account for time varying unobservable confounders. 

Third, the evaluation of the fit of the SCM model is very transparent since outcome values for the treated unit and its SCG should ideally be identical during the pre-treatment period \citep{abadie2010synthetic,abadie2021using, billmeier2013assessing}. 

Finally, the condition of non-negative weights summing up to one prevents extrapolation and simplifies interpretation of estimation results \citep{abadie2010synthetic,abadie2021using, billmeier2013assessing}.

Despite the numerous advantages of the SCM, \cite{abadie2021using} identifies a number of contextual and data requirements that need to be considered when applying SCM. First, SCM should not be applied if the treated unit is an outlier. This would for occur if a outcome or predictor value would lie outside the range of all potential control variables. Second, high volatility of outcomes can negatively affect the detectability of the treatment effect. Third, units that suffered from large idiosyncratic shocks should be eliminated from the donor pool. This would be the case if for example solar panels were destroyed by a storm or if a large-scale solar project was realized in one of the municipalities, resulting in a massive change in the outcome variable installed PV capacity per capita. Fourth, the validity of the non-interference assumption might be violated in the presence of spill-over effects. The treatment affect might be attenuated if the treatment status of unit \textit{i}, would affect geographically proximate areas. Fifth, the treatment should not be foreseeable by policy makers. Finally, since the SCG is a weighted sample of control units, substantial differences between its members might be averaged away, reducing the representativity of the treated unit. Whether this is the case can be checked by comparing the predictor balance of the SCG members.

Apart from contextual requirements, the application of synthetic controls requires aggregate data. Micro-level data can be used to obtain outcome or predictor values on the desired aggregate level. Moreover, pre- and post-intervention periods must be sufficiently long. In case that pre-treatment periods are too short, there exists the risk of over-fitting, while limited availability of post-intervention data might not capture the treatment effect to its full extent.

\newpage
\section{Results}

\begin{figure}[htb!]
    \centering
    \caption{Border 1, 2016-2022 (kW per capita)}
    \includegraphics[scale=0.2]{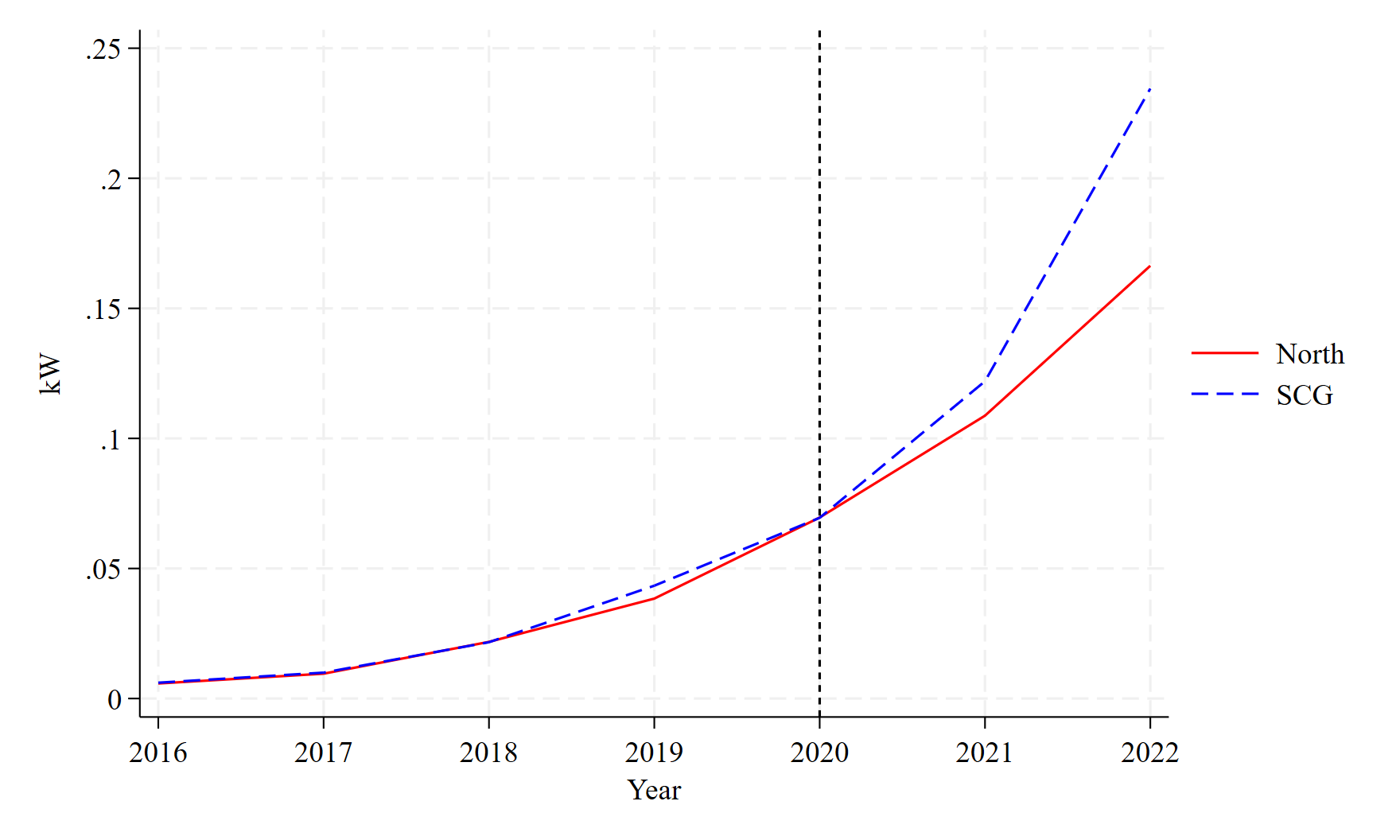}
    \label{fig:border_1}
\end{figure}

\begin{figure}[htb!]
    \centering
    \caption{Border 2, 2016-2022 (kW per capita)}
    \includegraphics[scale=0.2]{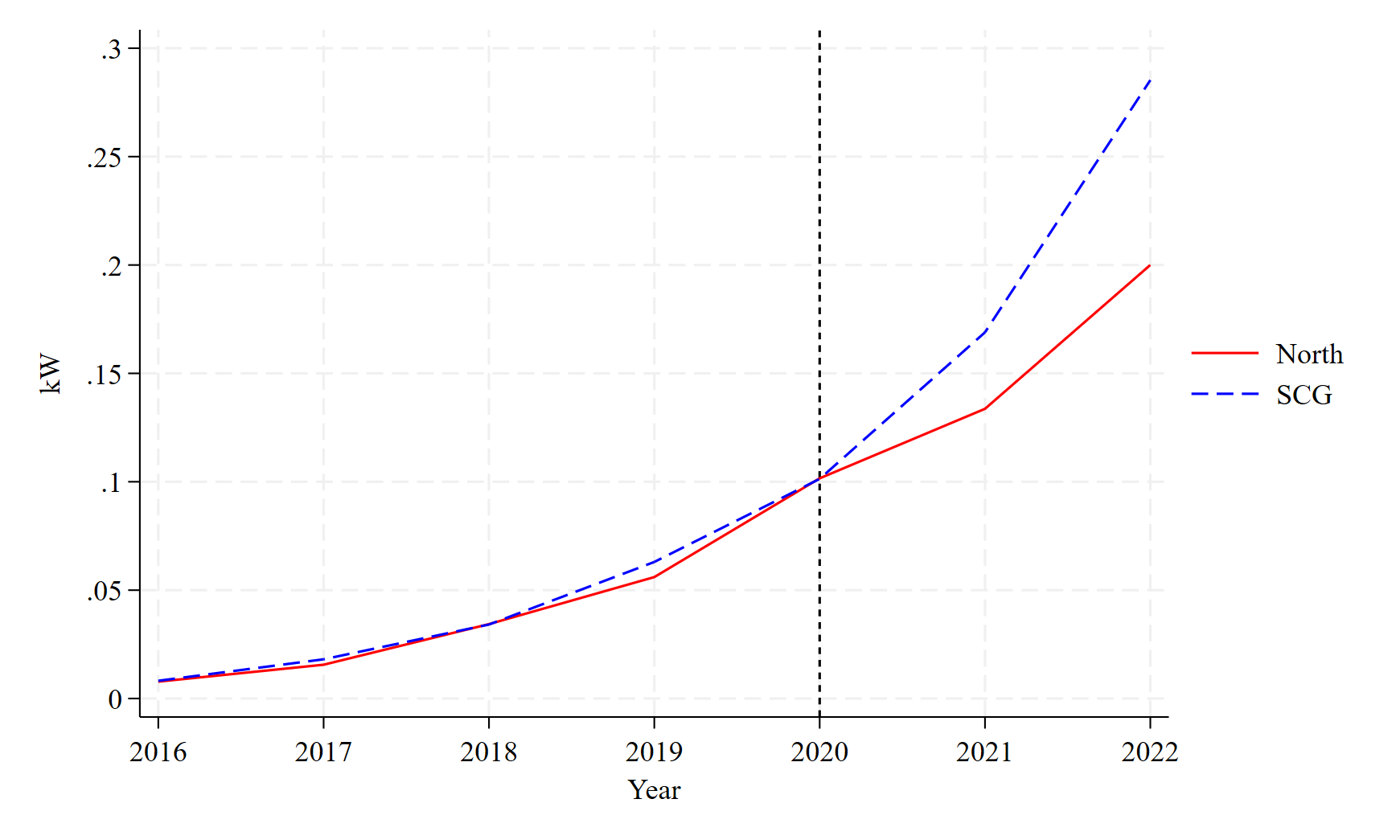}
    \label{fig:border_2}
\end{figure}

\begin{figure}[htb!]
    \centering
    \caption{Border 3, 2016-2022 (kW per capita)}
    \includegraphics[scale=0.2]{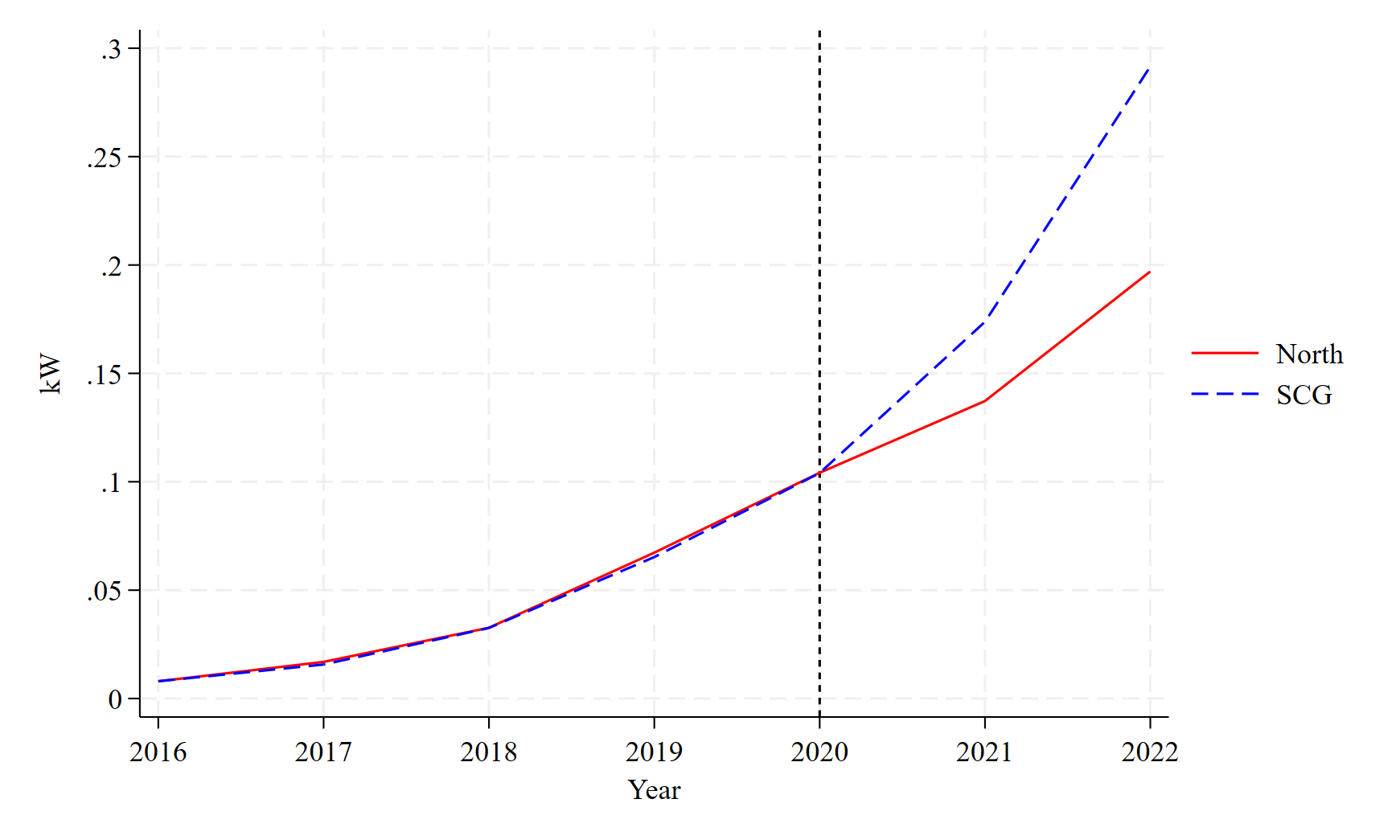}
    \label{fig:border_3}
\end{figure}

This section presents the regression results from the SCM estimations separately for the three definitions of the BZ border region. Installed PV capacity for the the border 1-3 and its SCGs are illustrated in figure \ref{fig:border_1}, \ref{fig:border_2}, and \ref{fig:border_3}, respectively. The SCG composition SCG is documented in table \ref{tab: scg_controlgroup}.

\begin{table}[htb!]
\centering
\caption{Synthetic control group composition}
\begin{tabular}{lrllrllr}
\hline 
\multicolumn{2}{c}{Border   1}               &  & \multicolumn{2}{c}{Border 2} &  & \multicolumn{2}{c}{Border 3} \\ \hline
Sandviken & 47\%                 &  & Gagnef     & 38\% &  & Gagnef      & 36\% \\
Gagnef     & 37\%                 &  & Vansbro   & 27\% &  & Vansbro     & 30\% \\
Borlänge    & 10\%                 &  & Norberg & 19\% &  & Borlänge    & 12\% \\
Vansbro     & 6\%                  &  & Borlänge    & 12\% &  & Falun       & 11\% \\
                      & \multicolumn{1}{l}{} &  & Sandviken & 4\%  &  & Norberg & 7\%  \\
                      & \multicolumn{1}{l}{} &  &                       &      &  & Leksand     & 4\%  \\ \hline
\end{tabular}
\label{tab: scg_controlgroup}
\end{table}

The municipalities used to construct the control groups are all located in geographical proximity of the BZ border between SE2 and SE3. All SCGs are sparse since they consist of a maximum of six out of the 234 municipalities included in the donor pool. Three municipalities are present in all SCGs, which can be accounted to the fact that border region contain by definition the same municipalities located in close proximities of the BZ border.

Figures \ref{fig:border_1}, \ref{fig:border_2}, and \ref{fig:border_3} reveal that outcome levels for the border and control regions are closely fitted during the pre-treatment period. A sizable deviation in outcome levels can only be observed in 2019 for border region 1 and 2, amounting to 12.9\% and 12.4\% in relative terms.

After 2020 PV uptake in the low-tariff border regions diverges clearly from its SCG. The size of the absolute and relative treatment effect are presented in table \ref{tab:scg_effect size}. The effect of electric tariff divergence arises already in 2021, but remains with 0.01 kW or 12.2\% insignificant for border 1, as deviations in pre-treatment levels are similar to the size of the 2021-estimate. For border 2 and 3 the effect size amounts to 0.04 kW or 22.4-22.6\% a single year after the emergence of electric tariff divergence in Sweden. A statistically and economically sizeable effect.

This situation changes when the data from 2022 is included. Two year after the start of treatment, the total effect rises to 0.07 kW and 0.09 kW, for border region 1 and both border regions 2 and 3, respectively. The size of relative effect size exceeds 40\% for all border regions but is with 48\% largest for border 3.

\begin{table}[htb!]
\centering
\caption{Absolute and relative treatment effect}
\begin{tabular}{lccc}
\hline

                        & Border 1 & Border 2 & Border 3 \\
\hline

Absolute  (kW)          &          &          &          \\
2020                    &  0.00    &  0.00    &  0.00    \\
2021                    &  -0.01   &  -0.04   & -0.04    \\
2022                    & -0.07    &  -0.09   & -0.09    \\ \hline
Relative   (\%)         &          &          &          \\
2020                    &  0.02    &  0.02    &  0.02    \\
2021                    &  -12.2   &  -26.4   &  -26.6   \\
2022                    & -40.9    & -42.6    & -48.0    \\ \hline
\end{tabular}
\label{tab:scg_effect size}
\end{table}

Important when evaluating the representativity of the SCG, is to compare the predictor balance of the SCG to the treated unit. These figures are compared in table \ref{tab:predictor balance} in absolute terms. Differences are most pronounced in radiation variation and increase when the border definition is widened. Differences in radiation variation are greatest for border 3 and its SCG amounting to a factor of 15 or 34.4\%. Nonetheless, this is unavoidable due to the geographic location of the BZ border. It is not considered as a serious issue here since differences in average radiation levels are far less pronounces amounting to a maximum of 46 kWh per \(m^2\) or 4.1\%.

\begin{table}[!ht]
    \centering
    \caption{Predictor balance}
    \begin{tabular}{|l|cc|cc|cc|}
    \hline
        & \multicolumn{2}{c|}{Border 1} & \multicolumn{2}{c|}{Border 2} & \multicolumn{2}{c|}{Border 3}  \\ \hline
        &      &        &       &          &          &          \\
         & Treated & SCG & Treated & SCG & Treated & SCG\\ \hline
        &      &        &       &          &          &          \\
        Income & 194063 & 193843 & 194429 & 194750 & 195729 & 196843 \\
        &      &        &       &          &          &          \\
        Radiation  & 1155 & 1184 & 1127 & 1164 & 1118 & 1164 \\
        &      &        &       &          &          &          \\
        Radiation variation & 36 & 28 & 39 & 28 & 43 & 28 \\
        &      &        &       &          &          &          \\
        Unemployment & 9\% & 8\% & 8\% & 8\% & 8\% & 7\% \\
        &      &        &       &          &          &          \\
        Small houses & 64\% & 67\% & 66\% & 72\% & 63\% & 73\% \\
        &      &        &       &          &          &          \\
        Capacity 2016 & 0.006 & 0.006 & 0.008 & 0.008 & 0.008 & 0.008 \\
        &      &        &       &          &          &          \\
        Capacity 2018 & 0.022 & 0.022 & 0.034 & 0.034 & 0.033 & 0.033 \\
        &      &        &       &          &          &          \\
        Capacity 2020 & 0.070 & 0.069 & 0.102 & 0.101 & 0.104 & 0.104 \\
        &      &        &       &          &          &          \\ \hline
    \end{tabular}
    \label{tab:predictor balance}
\end{table}

Apart from radiation variation, differences between the border region and their SCGs is greatest in terms of the unemployment rate. For border 2, the unemployment rate in the North exceeds that of the SCG by 0.51 percentage points which corresponds to a relative difference of 6.32\%.

\newpage
\section{Discussion}

This section discusses the implication of the estimation results for future PV expansion in Sweden and other European countries. Moreover, the use of SCM is critically evaluated and the results are compared to an alternative DiD estimation. Finally, the suitability of the alternative border region definitions is assessed to create a benchmark for future studies.

\subsection{Implications of the empirical findings}

The main finding of the study is that electricity price divergence since 2020 reduces PV uptake in the low-tariff BZ in Northern Sweden by 40.9\% to 48\% depending on the border region specification. This supports empirical findings that identify economic incentives as the main determinant of PV uptake in Sweden and elsewhere \citep{palm2020early,gautier2020pv,balta2015regional,borenstein2017private,assunccao2017developing}. Moreover, the study supports previous findings on the efficiency of BZs as an instrument to direct renewable energy investments towards areas with greatest imbalances in electricity supply and demand \citep{entsoe2023,hurta2022impact}.

Positive effects on the supply balance are contrasted by the rapid divergence in PV uptake across the country, which could harm countries prospects to realize their full PV potential and undergo a successful transition towards renewable generation. This is less concerning in Sweden with overall low PV uptake in an international comparison, a 70\% contribution of renewable sources to total electricity generation, and most of the low-tariff BZ being deemed technically unsuitable for an expansion of PV generation \citep{iea2022,castillo2016assessment,smhi2022}. Nonetheless, it should be considered in countries with low initial contribution of PV and overall renewable electricity generation such as the Czech Republic, PV investments would only be incentivized in the high-tariff BZ \citep{owd2023solarshare,owd2023electricity}. Similarly, existing inequalities in PV uptake would be reinforced in countries like Germany. Southern German states already display the highest levels of PV penetration and are expected to face the highest electric tariff rates if separate BZ are created. Meanwhile, incentives for PV investments in Northern Germany with the largest remaining potential would deteriorate along with the prospects for a successful German energy transition \citep{winter2019german,tagesschau2022,entsoe2023}.

Short-run imbalances in PV investments might become persistent in the long-run since their exist limitation to regional catch-up potential. Europe is currently facing a severe shortage of qualified solar installers. Despite massive demand, uninstalled PV modules are being stockpiled throughout the continent \citep{bloomberg2022shortage,blomwestergren2022brist}. Even if this issue was addressed, the problem remains that Europe imports the vast majority of its solar modules from China. Any kind of trade disruptions could therefore harm both, a successful energy transition and electricity supply security in Europe \citep{hancock2023}.

Furthermore, increased PV investments in high-tariff BZ could compromise grid stability due to the volatility of PV generation \citep{castillo2016assessment}. Increasing reliance on solar generation would require additional investment in grid infrastructure and storage capacity, which has resulted in multiple confrontations between locals and public authorities in the past \citep{hurta2022impact,schmidt2016optimal}. Local resistance has been so severe that governments in Sweden and Germany have discontinued funding and implemented laws impeding the realization of large scale energy projects to win over voters \citep{tidoavtalet2022,stede2019strikte}. 

Nonetheless, BZ creation might enact sufficient pressure on policy makers to raise community acceptance and remove legal impediments against energy infrastructure projects in general. According to \cite{hogan2022makes} and \cite{sena2016social} the majority of local resistance arises from the lack of financial returns from energy projects and their inability to influence the planning process. Both issues can be tackled by granting local communities full or co-ownership of the energy projects. Greater involvement of locals in the decision making process would be of special importance if separate BZ are created in since the Southern part of the country faces the biggest electricity shortage and is home to the fiercest opponents of energy projects.

\subsection{Is the use of synthetic controls justified?}

While the estimation results are economically and statistically significant, it must be evaluated whether the use of the SCM is appropriate for the present study. First of all, contextual aspects are assessed. The validity of this study would be compromised if policy makers were able to foresee the impact of price divergence on PV uptake when BZ were created in 2011 \citep{billmeier2013assessing}. This seems like an unlikely scenario since solar generation contributed back then less than 0.01\% to total electricity generation in Sweden \citep{owd2023solarshare}. Moreover, electricity price divergence occurred merely in 2020, one decade after BZ were first created \citep{energimyndigheten2022,vattenfall2022}. Finally, price divergence is partly driven by forces outside of Sweden namely the European electricity market integration and the absence of BZ in Germany combined with high demand, which incentivize exports even when electricity is in low supply in SE3 and SE4 \citep{dagensindustri2023tyskland}.

The temporal distance between investment decision and realization projects might be a challenge in this study, since data is only available two years after the start of treatment. Moreover, the lack of solar installers that affects the entire EU is also apparent in Sweden \citep{blomwestergren2022brist}. Waiting times for PV installations amount currently to about four month but vary across municipalities. The unequal access to PV installers could bias the estimation results in case that the access differs between the Northern BZ border and its SCG. The treatment effect might be overemphasized if access to solar installers was better among the SCG, while superior excess in the Northern BZ border would lead to its attenuation. 
 
Apart from contextual requirements, it must be ensured that the regression data is suited for the application of SCM. Therefore, a number of statistical tests have been performed. First, neither outcome nor control variables should lie outside the range of observations from the donor pool during the pre-treatment period. Moreover, results are most significant when outcomes in the post-treatment period lie outside the range of outcome levels of the donor pool. Whether this is the case can be accessed with help of a permutation distribution which displays the maximum and minimum levels of PV uptake among all municipalities included in the donor pool compared to the three border regions \citep{abadie2021using}.

\begin{figure}[htb!]
    \centering
    \caption{Permutation distribution, 2016-2022 (kW per capita)}
    \includegraphics[width=0.75\textwidth]{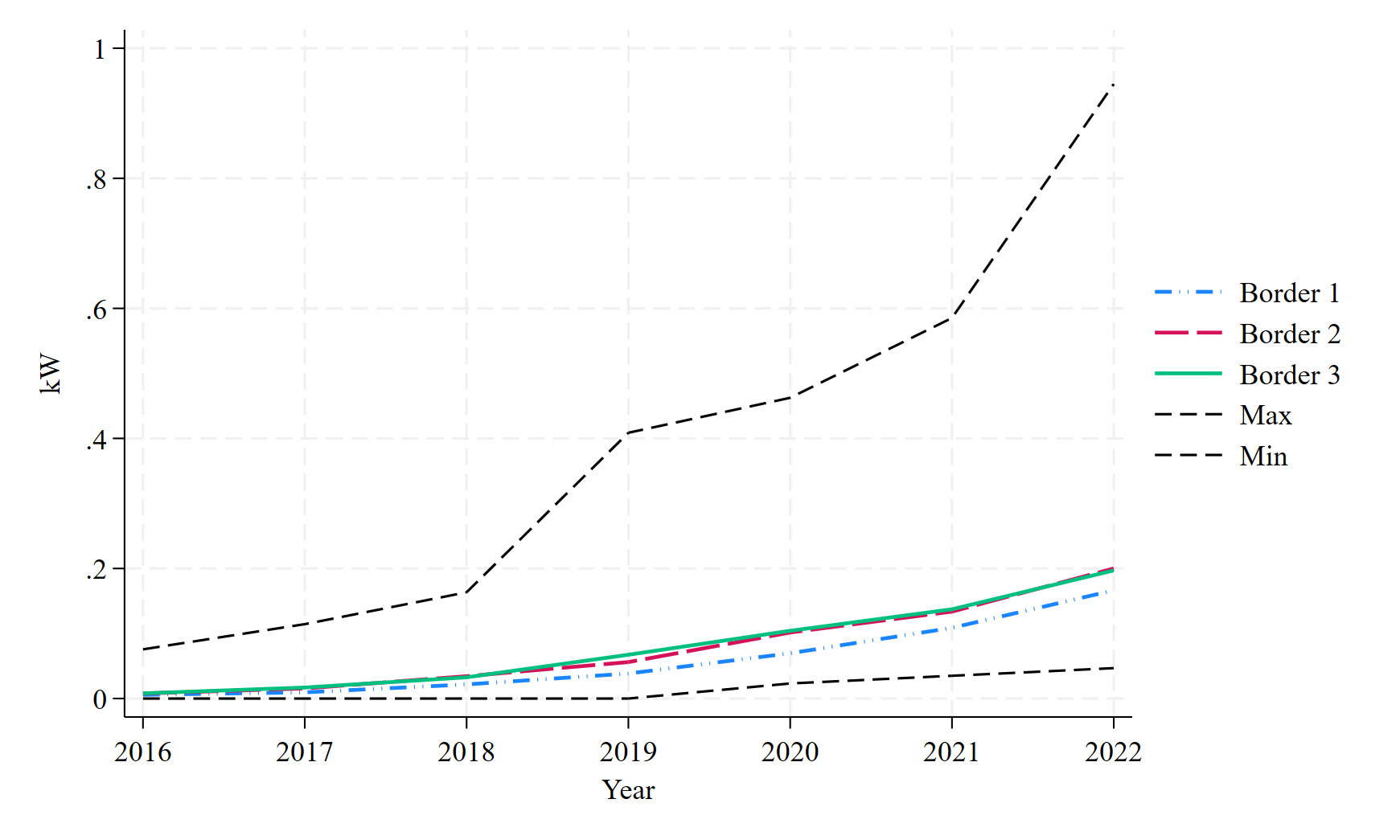}
    \label{fig:Permutation}
\end{figure}

The permutation of the outcome variable installed PV capacity per capita is illustrated in figure \ref{fig:Permutation}. The dashed lines indicate the maximum and minimum values from the donor pool. All border regions fulfill the condition that their outcome value are contained within the range of the donor pool. Nonetheless, the outcomes of the border regions remain within the range of outcomes in the high-price area casting some doubt on the significance of the estimation results, since there exist municipalities in the high-tariff areas with considerably lower levels of PV uptake.

Placebo tests are conducted to check for significant deviation in outcome levels between the border regions and their SCGs before actual treatment starts in 2020. For this purpose, the SCM regressions are repeated assuming the same placebo treatment period in 2018, with one less pre-treatment value in 2020. 

\begin{figure}[htb!]
    \centering
    \caption{SCG - Placebo test Border 1, 2016-2022 (kW)}
    \includegraphics[scale=0.23]{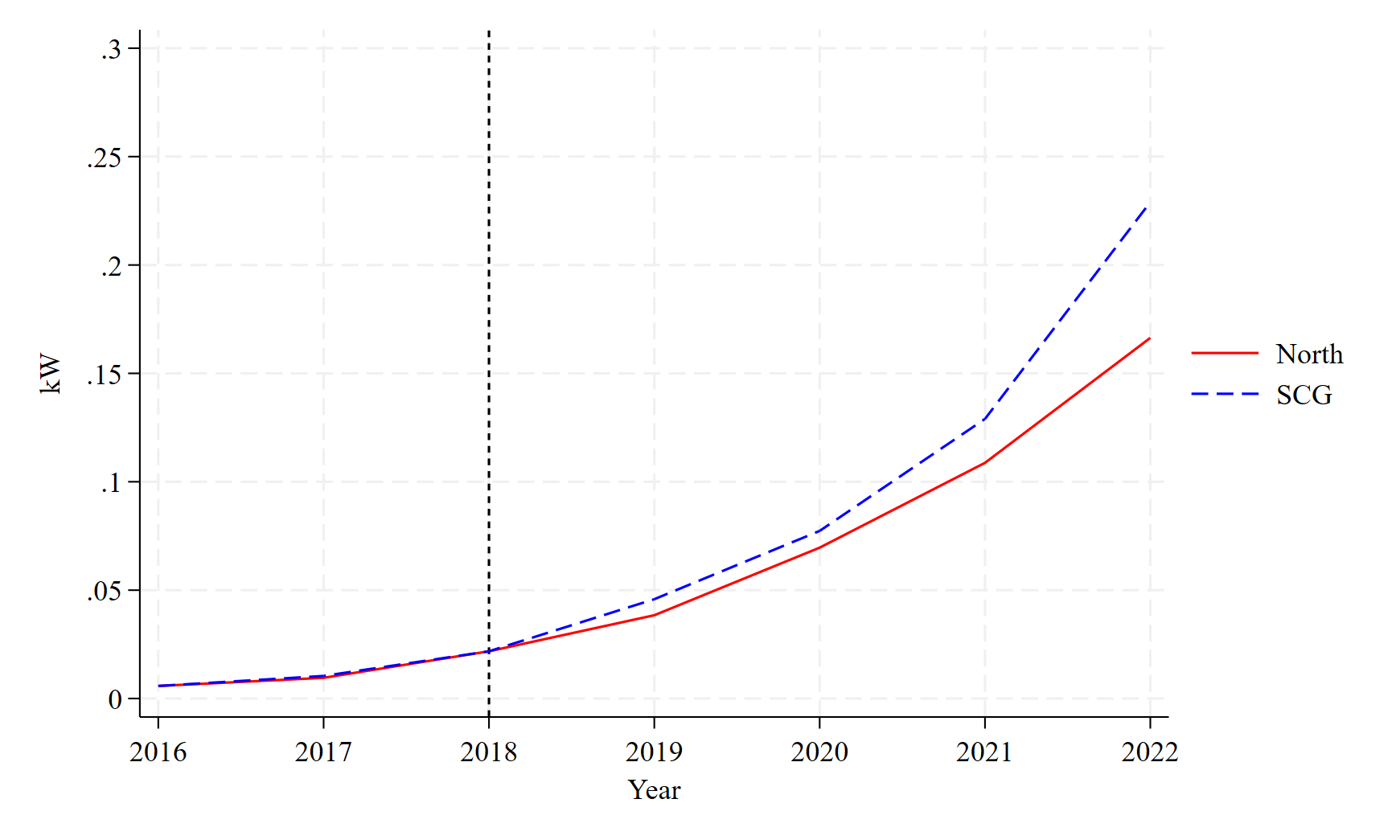}
    \label{fig:placebo_b1}
\end{figure}

\begin{figure}[htb!]
    \centering
    \caption{SCG - Placebo test Border 2, 2016-2022 (kW)}
    \includegraphics[scale=0.23]{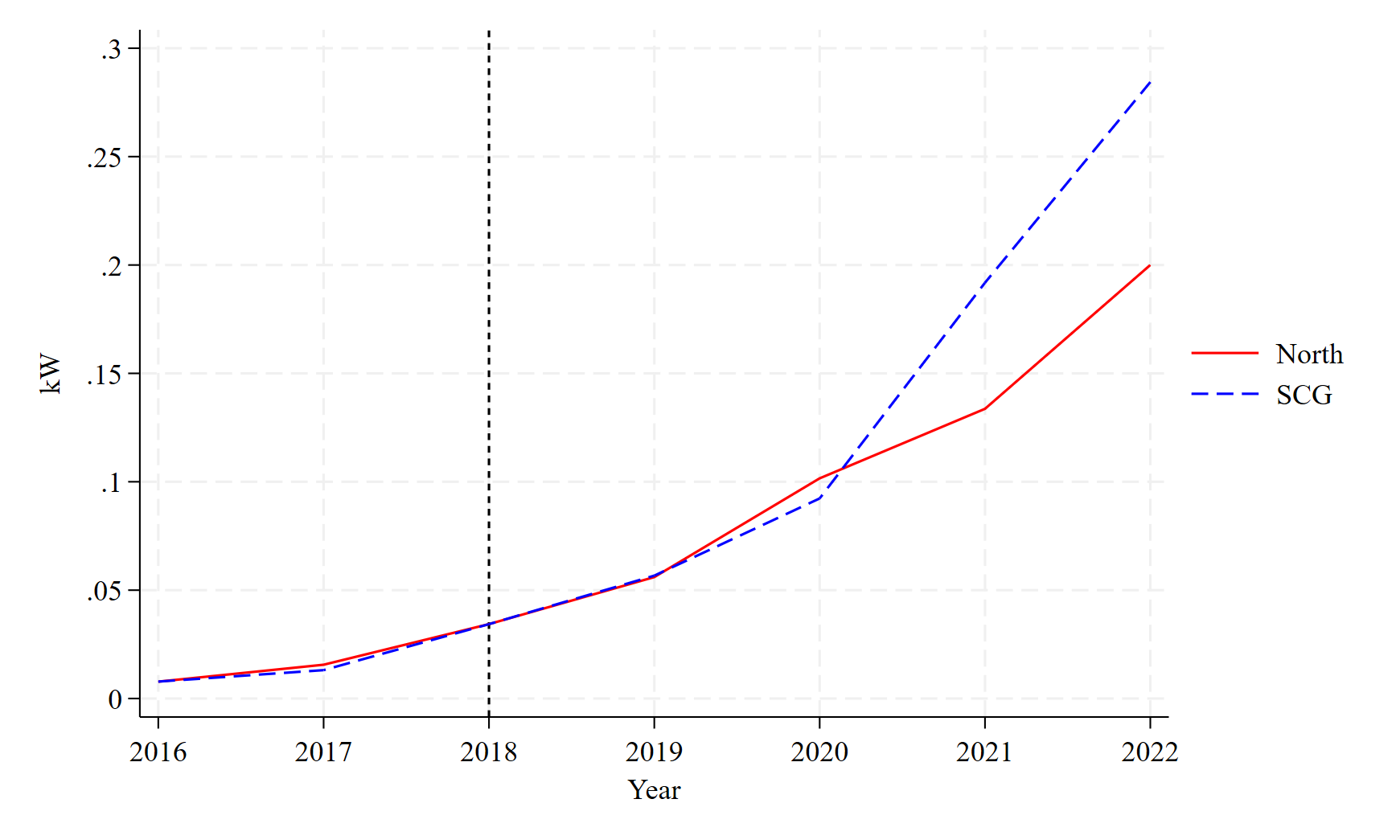}
    \label{fig:placebo_b2}
\end{figure}

\begin{figure}[htb!]
    \centering
    \caption{SCG - Placebo test Border 3, 2016-2022 (kW)}
    \includegraphics[scale=0.23]{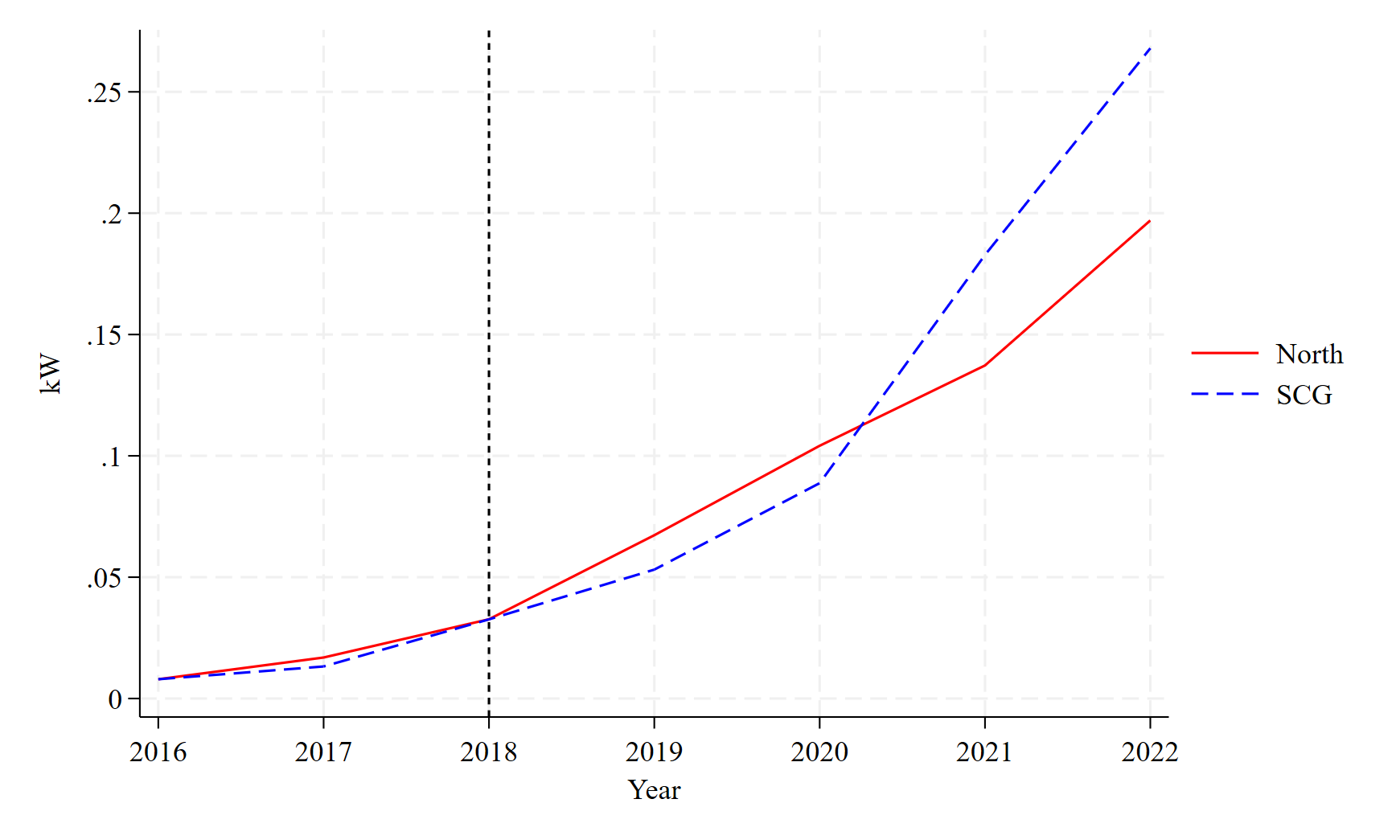}
    \label{fig:placebo_b3}
\end{figure}

The placebo test results are presented in figures \ref{fig:placebo_b1}, \ref{fig:placebo_b2}, and \ref{fig:placebo_b3}. Significant deviation between the Norther BZ border and its SCG arises only after 2020. This applies especially to border region 2 and 3, where the outcome of the treated unit exceeds that of its SCG. This supports the hypothesis that price divergence from 2022 onwards is the main driving force for deviations in PV uptake across BZ.

Finally, there exists a risk that the SCG does not properly represent the treatment unit. This problem would arise if there exist large variations in outcome and control variables between the members of the SCG, which would simply be \textit{averaged away} due to the choice of the estimation method \citep{abadie2021using}. To investigate this issue, values of outcome and control variables of individual members of the SCGs are presented in table \ref{tab:scg_variation}. All values of the control variables are the observations for 2016, while the value of the outcome variable is observed at three points in time, 2016, 2018, and 2020, which are also used to construct the SCGs.

Table \ref{tab:scg_variation} reveals that there exist significant variation within the SCGs regarding installed PV potential and the housing structure in the municipalities. For all SCGs the share of small houses varies by a factor of two or 45 percentage points. This has sere effects on incentives and potential for PV capacity which is especially visible in the of Borlänge, which is part of all SCGs and has both the lowest share of small houses and PV capacity installed.

\subsection{Alternative estimation}

The only other paper that uses electricity price divergence across BZ in Scandinavia as a natural experiment to estimate its effect on renewable energy technology investments was conducted by \cite{mauritzen2023great}, based on a DiD approach. The study's results are somewhat conflicting to ours since \citeauthor{mauritzen2023great} only detects an economically insignificant effect. While this might simply be driven by the different nature of the energy technologies or the country specific context, alternative DiD estimation of PV uptake in Sweden are presented in the appendix to investigate whether our results are robust to the choice of the estimation methods.

As presented in table \ref{tab:did_results}, the only statistically significant result are reported for border 3 when control variables are included. This effect is however very small in magnitude amounting to no more than 0.003 kW, corresponding to approximately 1.5\% of installed PV capacity in border 3 in 2022. 

Furthermore, DiD are deemed unsuitable to investigate PV uptake in Sweden since the underlying assumption that nothing but the electric tariff divergence should drive differences in PV uptake is violated. The results of  a placebo test presented in table \ref{tab:did_placebo} finds a significant \textit{average treatment effect} (ATET) when treatment is assumed to take place in 2018 instead of 2020. While the application of DiD might only be problematic in the Swedish context, future studies should carefully evaluate the validity of utilizing DiD to estimate the impact of price divergence across BZ on PV uptake.

\subsection{The choice of the border region}

Regarding the choice of a suitable border region, border 3 shows the best results when fitting the outcome data during the pre-treatment period. Moreover, the size of the estimated is largest among the alternative border regions.

The downside of selecting a wider border definition, are the growing imbalances in the predictor balance. In this study this concerns especially solar radiation levels since the low-tariff BZ is located in the North of the country and therefore further from the equator. To minimize underlying differences between the northern BZ border and its SCG, a narrow definition is preferred. Based on this considerations border 1 should be selected.

As a compromise between the two aspects border 2 is the preferred definition of the border region in this study. Under this definition the fit of the outcome graph is nearly as good as for border 3 (see figure \ref{fig:border_2}), while deviations in the predictor balance are considerable smaller (see table \ref{tab:predictor balance}).

\newpage
\section{Conclusion}

In the light of the EU's plan to divide several European countries into multiple BZ, the question arises whether BZ creation could lead to imbalances in PV uptake within countries. This study sheds light on this question by using electricity price divergence between Northern and Southern Swedish BZ after 2020 as a natural experiment to estimate the effect of electric-tariff divergence on unequal uptake of PV systems across BZ based on the SCM.

The results reveal that installed PV potential in low-tariff areas is 40.9-48\% less compared to high-tariff areas two years after the emergence of price differentials. These findings hold crucial implications for the success of the green transition of the European energy sector in Europe. Since PV uptake is mainly incentivized in high-tariff BZ, PV expansion might be limited to parts of the country or amplify existing imbalances. Not utilizing PV to its full potential entails the risk of failing to meet the EU's climate targets with respect to renewable electricity generation and carbon emission reduction.

Moreover, this study has made a methodological contribution by comparing the SCM to a DiD approach and different definitions of the BZ border when estimating the impact of electricity price divergence of PV uptake. The estimation results reveal, that DiD is not suitable in the Swedish case since the underlying assumptions are violated. Furthermore, the detected effects lack economical and statistical significance. 

Regarding the definition of the border region, border 2 is considered as the best choice since it combines a good fit of the outcome variable with a limited difference in the predictor variables.

Future studies in the field should aim to estimate the impact of electricity price divergence across BZ on PV investment in other countries or of other renewable energy sources within Sweden. The impact of electric tariff divergence on realized wind power projects in Sweden would be of special interest due to the technologies importance for electricity generation and the social conflict that arises around its diffusion \citep{bergek2010wind}.

\newpage
% \bibliography{Energy}

\newpage
\section*{Appendix}

\begin{longtable}{lllllll}
\caption{Municipality list}\\
\hline
        \textbf{Municipality} & \textbf{State (län)} & \textbf{Bidding zone} & \textbf{Border 1} & \textbf{Border 2} & \textbf{Border 3} \\ \hline
        \textbf{Arvidsjaur} & Norrbottens län & SE1 & 0 & 0 & 0 \\ 
        \textbf{Arjeplog} & Norrbottens län & SE1 & 0 & 0 & 0 \\ 
        \textbf{Jokkmokk} & Norrbottens län & SE1 & 0 & 0 & 0 \\ 
        \textbf{Överkalix} & Norrbottens län & SE1 & 0 & 0 & 0 \\ 
        \textbf{Kalix} & Norrbottens län & SE1 & 0 & 0 & 0 \\ 
        \textbf{Övertorneå} & Norrbottens län & SE1 & 0 & 0 & 0 \\ 
        \textbf{Pajala} & Norrbottens län & SE1 & 0 & 0 & 0 \\ 
        \textbf{Gällivare} & Norrbottens län & SE1 & 0 & 0 & 0 \\ 
        \textbf{Älvsbyn} & Norrbottens län & SE1 & 0 & 0 & 0 \\ 
        \textbf{Luleå} & Norrbottens län & SE1 & 0 & 0 & 0 \\ 
        \textbf{Piteå} & Norrbottens län & SE1 & 0 & 0 & 0 \\ 
        \textbf{Boden} & Norrbottens län & SE1 & 0 & 0 & 0 \\ 
        \textbf{Haparanda} & Norrbottens län & SE1 & 0 & 0 & 0 \\ 
        \textbf{Kiruna} & Norrbottens län & SE1 & 0 & 0 & 0 \\ 
        \textbf{Vindeln} & Västerbottens län & SE1 \& SE2 & 0 & 0 & 0 \\ 
        \textbf{Robertsfors} & Västerbottens län & SE1 \& SE2 & 0 & 0 & 0 \\ 
        \textbf{Norsjö} & Västerbottens län & SE1 \& SE2 & 0 & 0 & 0 \\ 
        \textbf{Malå} & Västerbottens län & SE1 \& SE2 & 0 & 0 & 0 \\ 
        \textbf{Sorsele} & Västerbottens län & SE1 \& SE2 & 0 & 0 & 0 \\ 
        \textbf{Umeå} & Västerbottens län & SE1 \& SE2 & 0 & 0 & 0 \\ 
        \textbf{Lycksele} & Västerbottens län & SE1 \& SE2 & 0 & 0 & 0 \\ 
        \textbf{Skellefteå} & Västerbottens län & SE1 \& SE2 & 0 & 0 & 0 \\ 
        \textbf{Nordanstig} & Gävleborgs län & SE2 & 0 & 0 & 1 \\ 
        \textbf{Ljusdal} & Gävleborgs län & SE2 & 1 & 1 & 1 \\ 
        \textbf{Söderhamn} & Gävleborgs län & SE2 & 1 & 1 & 1 \\ 
        \textbf{Bollnäs} & Gävleborgs län & SE2 & 1 & 1 & 1 \\ 
        \textbf{Hudiksvall} & Gävleborgs län & SE2 & 0 & 1 & 1 \\ 
        \textbf{Ånge} & Västernorrlands län & SE2 & 0 & 1 & 1 \\ 
        \textbf{Timrå} & Västernorrlands län & SE2 & 0 & 0 & 0 \\ 
        \textbf{Härnösand} & Västernorrlands län & SE2 & 0 & 0 & 0 \\ 
        \textbf{Sundsvall} & Västernorrlands län & SE2 & 0 & 0 & 1 \\ 
        \textbf{Kramfors} & Västernorrlands län & SE2 & 0 & 0 & 0 \\ 
        \textbf{Sollefteå} & Västernorrlands län & SE2 & 0 & 0 & 0 \\ 
        \textbf{Örnsköldsvik} & Västernorrlands län & SE2 & 0 & 0 & 0 \\ 
        \textbf{Ragunda} & Jämtlands län & SE2 & 0 & 0 & 0 \\ 
        \textbf{Bräcke} & Jämtlands län & SE2 & 0 & 0 & 1 \\ 
        \textbf{Krokom} & Jämtlands län & SE2 & 0 & 0 & 0 \\ 
        \textbf{Strömsund} & Jämtlands län & SE2 & 0 & 0 & 0 \\ 
        \textbf{Åre} & Jämtlands län & SE2 & 0 & 0 & 1 \\ 
        \textbf{Berg} & Jämtlands län & SE2 & 0 & 1 & 1 \\ 
        \textbf{Härjedalen} & Jämtlands län & SE2 & 1 & 1 & 1 \\ 
        \textbf{Östersund} & Jämtlands län & SE2 & 0 & 0 & 1 \\ 
        \textbf{Nordmaling} & Västerbottens län & SE2 & 0 & 0 & 0 \\ 
        \textbf{Bjurholm} & Västerbottens län & SE2 & 0 & 0 & 0 \\ 
        \textbf{Storuman} & Västerbottens län & SE2 & 0 & 0 & 0 \\ 
        \textbf{Dorotea} & Västerbottens län & SE2 & 0 & 0 & 0 \\ 
        \textbf{Vännäs} & Västerbottens län & SE2 & 0 & 0 & 0 \\ 
        \textbf{Vilhelmina} & Västerbottens län & SE2 & 0 & 0 & 0 \\ 
        \textbf{Åsele} & Västerbottens län & SE2 & 0 & 0 & 0 \\ 
        \textbf{Rättvik} & Dalarnas län & SE2 \& SE3 & 1 & 1 & 1 \\ 
        \textbf{Orsa} & Dalarnas län & SE2 \& SE3 & 1 & 1 & 1 \\ 
        \textbf{Älvdalen} & Dalarnas län & SE2 \& SE3 & 1 & 1 & 1 \\ 
        \textbf{Mora} & Dalarnas län & SE2 \& SE3 & 1 & 1 & 1 \\ 
        \textbf{Ockelbo} & Gävleborgs län & SE2 \& SE3 & 1 & 1 & 1 \\ 
        \textbf{Ovanåker} & Gävleborgs län & SE2 \& SE3 & 1 & 1 & 1 \\ 
        \textbf{Gävle} & Gävleborgs län & SE2 \& SE3 & 1 & 1 & 1 \\ 
        \textbf{Upplands Väsby} & Stockholms län & SE3 & 0 & 0 & 0 \\ 
        \textbf{Vallentuna} & Stockholms län & SE3 & 0 & 0 & 0 \\ 
        \textbf{Österåker} & Stockholms län & SE3 & 0 & 0 & 0 \\ 
        \textbf{Värmdö} & Stockholms län & SE3 & 0 & 0 & 0 \\ 
        \textbf{Järfälla} & Stockholms län & SE3 & 0 & 0 & 0 \\ 
        \textbf{Ekerö} & Stockholms län & SE3 & 0 & 0 & 0 \\ 
        \textbf{Huddinge} & Stockholms län & SE3 & 0 & 0 & 0 \\ 
        \textbf{Botkyrka} & Stockholms län & SE3 & 0 & 0 & 0 \\ 
        \textbf{Salem} & Stockholms län & SE3 & 0 & 0 & 0 \\ 
        \textbf{Haninge} & Stockholms län & SE3 & 0 & 0 & 0 \\ 
        \textbf{Tyresö} & Stockholms län & SE3 & 0 & 0 & 0 \\ 
        \textbf{Upplands-Bro} & Stockholms län & SE3 & 0 & 0 & 0 \\ 
        \textbf{Nykvarn} & Stockholms län & SE3 & 0 & 0 & 0 \\ 
        \textbf{Täby} & Stockholms län & SE3 & 0 & 0 & 0 \\ 
        \textbf{Danderyd} & Stockholms län & SE3 & 0 & 0 & 0 \\ 
        \textbf{Sollentuna} & Stockholms län & SE3 & 0 & 0 & 0 \\ 
        \textbf{Stockholm} & Stockholms län & SE3 & 0 & 0 & 0 \\ 
        \textbf{Södertälje} & Stockholms län & SE3 & 0 & 0 & 0 \\ 
        \textbf{Nacka} & Stockholms län & SE3 & 0 & 0 & 0 \\ 
        \textbf{Sundbyberg} & Stockholms län & SE3 & 0 & 0 & 0 \\ 
        \textbf{Solna} & Stockholms län & SE3 & 0 & 0 & 0 \\ 
        \textbf{Lidingö} & Stockholms län & SE3 & 0 & 0 & 0 \\ 
        \textbf{Vaxholm} & Stockholms län & SE3 & 0 & 0 & 0 \\ 
        \textbf{Norrtälje} & Stockholms län & SE3 & 0 & 0 & 1 \\ 
        \textbf{Sigtuna} & Stockholms län & SE3 & 0 & 0 & 0 \\ 
        \textbf{Nynäshamn} & Stockholms län & SE3 & 0 & 0 & 0 \\ 
        \textbf{Håbo} & Uppsala län & SE3 & 0 & 0 & 1 \\ 
        \textbf{Älvkarleby} & Uppsala län & SE3 & 1 & 1 & 1 \\ 
        \textbf{Knivsta} & Uppsala län & SE3 & 0 & 0 & 1 \\ 
        \textbf{Heby} & Uppsala län & SE3 & 1 & 1 & 1 \\ 
        \textbf{Tierp} & Uppsala län & SE3 & 1 & 1 & 1 \\ 
        \textbf{Uppsala} & Uppsala län & SE3 & 0 & 1 & 1 \\ 
        \textbf{Enköping} & Uppsala län & SE3 & 0 & 1 & 1 \\ 
        \textbf{Östhammar} & Uppsala län & SE3 & 0 & 1 & 1 \\ 
        \textbf{Vingåker} & Södermanlands län & SE3 & 0 & 0 & 0 \\ 
        \textbf{Gnesta} & Södermanlands län & SE3 & 0 & 0 & 0 \\ 
        \textbf{Nyköping} & Södermanlands län & SE3 & 0 & 0 & 0 \\ 
        \textbf{Oxelösund} & Södermanlands län & SE3 & 0 & 0 & 0 \\ 
        \textbf{Flen} & Södermanlands län & SE3 & 0 & 0 & 0 \\ 
        \textbf{Katrineholm} & Södermanlands län & SE3 & 0 & 0 & 0 \\ 
        \textbf{Eskilstuna} & Södermanlands län & SE3 & 0 & 0 & 0 \\ 
        \textbf{Strängnäs} & Södermanlands län & SE3 & 0 & 0 & 0 \\ 
        \textbf{Trosa} & Södermanlands län & SE3 & 0 & 0 & 0 \\ 
        \textbf{Ödeshög} & Östergötlands län & SE3 & 0 & 0 & 0 \\ 
        \textbf{Ydre} & Östergötlands län & SE3 & 0 & 0 & 0 \\ 
        \textbf{Kinda} & Östergötlands län & SE3 & 0 & 0 & 0 \\ 
        \textbf{Boxholm} & Östergötlands län & SE3 & 0 & 0 & 0 \\ 
        \textbf{Åtvidaberg} & Östergötlands län & SE3 & 0 & 0 & 0 \\ 
        \textbf{Finspång} & Östergötlands län & SE3 & 0 & 0 & 0 \\ 
        \textbf{Valdemarsvik} & Östergötlands län & SE3 & 0 & 0 & 0 \\ 
        \textbf{Linköping} & Östergötlands län & SE3 & 0 & 0 & 0 \\ 
        \textbf{Norrköping} & Östergötlands län & SE3 & 0 & 0 & 0 \\ 
        \textbf{Söderköping} & Östergötlands län & SE3 & 0 & 0 & 0 \\ 
        \textbf{Motala} & Östergötlands län & SE3 & 0 & 0 & 0 \\ 
        \textbf{Vadstena} & Östergötlands län & SE3 & 0 & 0 & 0 \\ 
        \textbf{Mjölby} & Östergötlands län & SE3 & 0 & 0 & 0 \\ 
        \textbf{Aneby} & Jönköpings län & SE3 & 0 & 0 & 0 \\ 
        \textbf{Mullsjö} & Jönköpings län & SE3 & 0 & 0 & 0 \\ 
        \textbf{Habo} & Jönköpings län & SE3 & 0 & 0 & 0 \\ 
        \textbf{Jönköping} & Jönköpings län & SE3 & 0 & 0 & 0 \\ 
        \textbf{Nässjö} & Jönköpings län & SE3 & 0 & 0 & 0 \\ 
        \textbf{Eksjö} & Jönköpings län & SE3 & 0 & 0 & 0 \\ 
        \textbf{Tranås} & Jönköpings län & SE3 & 0 & 0 & 0 \\ 
        \textbf{Västervik} & Kalmar län & SE3 & 0 & 0 & 0 \\ 
        \textbf{Vimmerby} & Kalmar län & SE3 & 0 & 0 & 0 \\ 
        \textbf{Gotland} & Gotlands län & SE3 & 0 & 0 & 0 \\ 
        \textbf{Kungsbacka} & Hallands län & SE3 & 0 & 0 & 0 \\ 
        \textbf{Härryda} & Västra Götalands län & SE3 & 0 & 0 & 0 \\ 
        \textbf{Partille} & Västra Götalands län & SE3 & 0 & 0 & 0 \\ 
        \textbf{Öckerö} & Västra Götalands län & SE3 & 0 & 0 & 0 \\ 
        \textbf{Stenungsund} & Västra Götalands län & SE3 & 0 & 0 & 0 \\ 
        \textbf{Tjörn} & Västra Götalands län & SE3 & 0 & 0 & 0 \\ 
        \textbf{Orust} & Västra Götalands län & SE3 & 0 & 0 & 0 \\ 
        \textbf{Sotenäs} & Västra Götalands län & SE3 & 0 & 0 & 0 \\ 
        \textbf{Munkedal} & Västra Götalands län & SE3 & 0 & 0 & 0 \\ 
        \textbf{Tanum} & Västra Götalands län & SE3 & 0 & 0 & 0 \\ 
        \textbf{Dals-Ed} & Västra Götalands län & SE3 & 0 & 0 & 0 \\ 
        \textbf{Färgelanda} & Västra Götalands län & SE3 & 0 & 0 & 0 \\ 
        \textbf{Ale} & Västra Götalands län & SE3 & 0 & 0 & 0 \\ 
        \textbf{Lerum} & Västra Götalands län & SE3 & 0 & 0 & 0 \\ 
        \textbf{Vårgårda} & Västra Götalands län & SE3 & 0 & 0 & 0 \\ 
        \textbf{Bollebygd} & Västra Götalands län & SE3 & 0 & 0 & 0 \\ 
        \textbf{Grästorp} & Västra Götalands län & SE3 & 0 & 0 & 0 \\ 
        \textbf{Essunga} & Västra Götalands län & SE3 & 0 & 0 & 0 \\ 
        \textbf{Karlsborg} & Västra Götalands län & SE3 & 0 & 0 & 0 \\ 
        \textbf{Gullspång} & Västra Götalands län & SE3 & 0 & 0 & 0 \\ 
        \textbf{Tranemo} & Västra Götalands län & SE3 & 0 & 0 & 0 \\ 
        \textbf{Bengtsfors} & Västra Götalands län & SE3 & 0 & 0 & 0 \\ 
        \textbf{Mellerud} & Västra Götalands län & SE3 & 0 & 0 & 0 \\ 
        \textbf{Lilla Edet} & Västra Götalands län & SE3 & 0 & 0 & 0 \\ 
        \textbf{Mark} & Västra Götalands län & SE3 & 0 & 0 & 0 \\ 
        \textbf{Herrljunga} & Västra Götalands län & SE3 & 0 & 0 & 0 \\ 
        \textbf{Vara} & Västra Götalands län & SE3 & 0 & 0 & 0 \\ 
        \textbf{Götene} & Västra Götalands län & SE3 & 0 & 0 & 0 \\ 
        \textbf{Tibro} & Västra Götalands län & SE3 & 0 & 0 & 0 \\ 
        \textbf{Töreboda} & Västra Götalands län & SE3 & 0 & 0 & 0 \\ 
        \textbf{Göteborg} & Västra Götalands län & SE3 & 0 & 0 & 0 \\ 
        \textbf{Mölndal} & Västra Götalands län & SE3 & 0 & 0 & 0 \\ 
        \textbf{Kungälv} & Västra Götalands län & SE3 & 0 & 0 & 0 \\ 
        \textbf{Lysekil} & Västra Götalands län & SE3 & 0 & 0 & 0 \\ 
        \textbf{Uddevalla} & Västra Götalands län & SE3 & 0 & 0 & 0 \\ 
        \textbf{Strömstad} & Västra Götalands län & SE3 & 0 & 0 & 0 \\ 
        \textbf{Vänersborg} & Västra Götalands län & SE3 & 0 & 0 & 0 \\ 
        \textbf{Trollhättan} & Västra Götalands län & SE3 & 0 & 0 & 0 \\ 
        \textbf{Alingsås} & Västra Götalands län & SE3 & 0 & 0 & 0 \\ 
        \textbf{Borås} & Västra Götalands län & SE3 & 0 & 0 & 0 \\ 
        \textbf{Ulricehamn} & Västra Götalands län & SE3 & 0 & 0 & 0 \\ 
        \textbf{Åmål} & Västra Götalands län & SE3 & 0 & 0 & 0 \\ 
        \textbf{Mariestad} & Västra Götalands län & SE3 & 0 & 0 & 0 \\ 
        \textbf{Lidköping} & Västra Götalands län & SE3 & 0 & 0 & 0 \\ 
        \textbf{Skara} & Västra Götalands län & SE3 & 0 & 0 & 0 \\ 
        \textbf{Skövde} & Västra Götalands län & SE3 & 0 & 0 & 0 \\ 
        \textbf{Hjo} & Västra Götalands län & SE3 & 0 & 0 & 0 \\ 
        \textbf{Tidaholm} & Västra Götalands län & SE3 & 0 & 0 & 0 \\ 
        \textbf{Falköping} & Västra Götalands län & SE3 & 0 & 0 & 0 \\ 
        \textbf{Kil} & Värmlands län & SE3 & 0 & 0 & 0 \\ 
        \textbf{Eda} & Värmlands län & SE3 & 0 & 0 & 0 \\ 
        \textbf{Torsby} & Värmlands län & SE3 & 0 & 1 & 1 \\ 
        \textbf{Storfors} & Värmlands län & SE3 & 0 & 0 & 1 \\ 
        \textbf{Hammarö} & Värmlands län & SE3 & 0 & 0 & 0 \\ 
        \textbf{Munkfors} & Värmlands län & SE3 & 0 & 0 & 1 \\ 
        \textbf{Forshaga} & Värmlands län & SE3 & 0 & 0 & 1 \\ 
        \textbf{Grums} & Värmlands län & SE3 & 0 & 0 & 0 \\ 
        \textbf{Årjäng} & Värmlands län & SE3 & 0 & 0 & 0 \\ 
        \textbf{Sunne} & Värmlands län & SE3 & 0 & 0 & 1 \\ 
        \textbf{Karlstad} & Värmlands län & SE3 & 0 & 0 & 1 \\ 
        \textbf{Kristinehamn} & Värmlands län & SE3 & 0 & 0 & 0 \\ 
        \textbf{Filipstad} & Värmlands län & SE3 & 0 & 1 & 1 \\ 
        \textbf{Hagfors} & Värmlands län & SE3 & 0 & 1 & 1 \\ 
        \textbf{Arvika} & Värmlands län & SE3 & 0 & 0 & 1 \\ 
        \textbf{Säffle} & Värmlands län & SE3 & 0 & 0 & 0 \\ 
        \textbf{Lekeberg} & Örebro län & SE3 & 0 & 0 & 0 \\ 
        \textbf{Laxå} & Örebro län & SE3 & 0 & 0 & 0 \\ 
        \textbf{Hallsberg} & Örebro län & SE3 & 0 & 0 & 0 \\ 
        \textbf{Degerfors} & Örebro län & SE3 & 0 & 0 & 0 \\ 
        \textbf{Hällefors} & Örebro län & SE3 & 0 & 0 & 1 \\ 
        \textbf{Ljusnarsberg} & Örebro län & SE3 & 0 & 0 & 1 \\ 
        \textbf{Örebro} & Örebro län & SE3 & 0 & 0 & 0 \\ 
        \textbf{Kumla} & Örebro län & SE3 & 0 & 0 & 0 \\ 
        \textbf{Askersund} & Örebro län & SE3 & 0 & 0 & 0 \\ 
        \textbf{Karlskoga} & Örebro län & SE3 & 0 & 0 & 0 \\ 
        \textbf{Nora} & Örebro län & SE3 & 0 & 0 & 0 \\ 
        \textbf{Lindesberg} & Örebro län & SE3 & 0 & 0 & 0 \\ 
        \textbf{Skinnskatteberg} & Västmanlands län & SE3 & 0 & 0 & 0 \\ 
        \textbf{Surahammar} & Västmanlands län & SE3 & 0 & 0 & 1 \\ 
        \textbf{Kungsör} & Västmanlands län & SE3 & 0 & 0 & 0 \\ 
        \textbf{Hallstahammar} & Västmanlands län & SE3 & 0 & 0 & 0 \\ 
        \textbf{Norberg} & Västmanlands län & SE3 & 0 & 0 & 1 \\ 
        \textbf{Västerås} & Västmanlands län & SE3 & 0 & 0 & 1 \\ 
        \textbf{Sala} & Västmanlands län & SE3 & 0 & 1 & 1 \\ 
        \textbf{Fagersta} & Västmanlands län & SE3 & 0 & 0 & 1 \\ 
        \textbf{Köping} & Västmanlands län & SE3 & 0 & 0 & 0 \\ 
        \textbf{Arboga} & Västmanlands län & SE3 & 0 & 0 & 0 \\ 
        \textbf{Vansbro} & Dalarnas län & SE3 & 1 & 1 & 1 \\ 
        \textbf{Malung-Sälen} & Dalarnas län & SE3 & 1 & 1 & 1 \\ 
        \textbf{Gagnef} & Dalarnas län & SE3 & 0 & 1 & 1 \\ 
        \textbf{Leksand} & Dalarnas län & SE3 & 1 & 1 & 1 \\ 
        \textbf{Smedjebacken} & Dalarnas län & SE3 & 0 & 0 & 1 \\ 
        \textbf{Falun} & Dalarnas län & SE3 & 1 & 1 & 1 \\ 
        \textbf{Borlänge} & Dalarnas län & SE3 & 0 & 1 & 1 \\ 
        \textbf{Säter} & Dalarnas län & SE3 & 0 & 1 & 1 \\ 
        \textbf{Hedemora} & Dalarnas län & SE3 & 0 & 1 & 1 \\ 
        \textbf{Avesta} & Dalarnas län & SE3 & 0 & 1 & 1 \\ 
        \textbf{Ludvika} & Dalarnas län & SE3 & 0 & 1 & 1 \\ 
        \textbf{Hofors} & Gävleborgs län & SE3 & 1 & 1 & 1 \\ 
        \textbf{Sandviken} & Gävleborgs län & SE3 & 1 & 1 & 1 \\ 
        \textbf{Gnosjö} & Jönköpings län & SE3 \& SE4 & 0 & 0 & 0 \\ 
        \textbf{Gislaved} & Jönköpings län & SE3 \& SE4 & 0 & 0 & 0 \\ 
        \textbf{Vaggeryd} & Jönköpings län & SE3 \& SE4 & 0 & 0 & 0 \\ 
        \textbf{Sävsjö} & Jönköpings län & SE3 \& SE4 & 0 & 0 & 0 \\ 
        \textbf{Vetlanda} & Jönköpings län & SE3 \& SE4 & 0 & 0 & 0 \\ 
        \textbf{Växjö} & Kronobergs län & SE3 \& SE4 & 0 & 0 & 0 \\ 
        \textbf{Hultsfred} & Kalmar län & SE3 \& SE4 & 0 & 0 & 0 \\ 
        \textbf{Oskarshamn} & Kalmar län & SE3 \& SE4 & 0 & 0 & 0 \\ 
        \textbf{Varberg} & Hallands län & SE3 \& SE4 & 0 & 0 & 0 \\ 
        \textbf{Svenljunga} & Västra Götalands län & SE3 \& SE4 & 0 & 0 & 0 \\ 
        \textbf{Värnamo} & Jönköpings län & SE4 & 0 & 0 & 0 \\ 
        \textbf{Uppvidinge} & Kronobergs län & SE4 & 0 & 0 & 0 \\ 
        \textbf{Lessebo} & Kronobergs län & SE4 & 0 & 0 & 0 \\ 
        \textbf{Tingsryd} & Kronobergs län & SE4 & 0 & 0 & 0 \\ 
        \textbf{Alvesta} & Kronobergs län & SE4 & 0 & 0 & 0 \\ 
        \textbf{Älmhult} & Kronobergs län & SE4 & 0 & 0 & 0 \\ 
        \textbf{Markaryd} & Kronobergs län & SE4 & 0 & 0 & 0 \\ 
        \textbf{Ljungby} & Kronobergs län & SE4 & 0 & 0 & 0 \\ 
        \textbf{Högsby} & Kalmar län & SE4 & 0 & 0 & 0 \\ 
        \textbf{Torsås} & Kalmar län & SE4 & 0 & 0 & 0 \\ 
        \textbf{Mörbylånga} & Kalmar län & SE4 & 0 & 0 & 0 \\ 
        \textbf{Mönsterås} & Kalmar län & SE4 & 0 & 0 & 0 \\ 
        \textbf{Emmaboda} & Kalmar län & SE4 & 0 & 0 & 0 \\ 
        \textbf{Kalmar} & Kalmar län & SE4 & 0 & 0 & 0 \\ 
        \textbf{Nybro} & Kalmar län & SE4 & 0 & 0 & 0 \\ 
        \textbf{Borgholm} & Kalmar län & SE4 & 0 & 0 & 0 \\ 
        \textbf{Olofström} & Blekinge län & SE4 & 0 & 0 & 0 \\ 
        \textbf{Karlskrona} & Blekinge län & SE4 & 0 & 0 & 0 \\ 
        \textbf{Ronneby} & Blekinge län & SE4 & 0 & 0 & 0 \\ 
        \textbf{Karlshamn} & Blekinge län & SE4 & 0 & 0 & 0 \\ 
        \textbf{Sölvesborg} & Blekinge län & SE4 & 0 & 0 & 0 \\ 
        \textbf{Svalöv} & Skåne län & SE4 & 0 & 0 & 0 \\ 
        \textbf{Staffanstorp} & Skåne län & SE4 & 0 & 0 & 0 \\ 
        \textbf{Burlöv} & Skåne län & SE4 & 0 & 0 & 0 \\ 
        \textbf{Vellinge} & Skåne län & SE4 & 0 & 0 & 0 \\ 
        \textbf{Östra Göinge} & Skåne län & SE4 & 0 & 0 & 0 \\ 
        \textbf{Örkelljunga} & Skåne län & SE4 & 0 & 0 & 0 \\ 
        \textbf{Bjuv} & Skåne län & SE4 & 0 & 0 & 0 \\ 
        \textbf{Kävlinge} & Skåne län & SE4 & 0 & 0 & 0 \\ 
        \textbf{Lomma} & Skåne län & SE4 & 0 & 0 & 0 \\ 
        \textbf{Svedala} & Skåne län & SE4 & 0 & 0 & 0 \\ 
        \textbf{Skurup} & Skåne län & SE4 & 0 & 0 & 0 \\ 
        \textbf{Sjöbo} & Skåne län & SE4 & 0 & 0 & 0 \\ 
        \textbf{Hörby} & Skåne län & SE4 & 0 & 0 & 0 \\ 
        \textbf{Höör} & Skåne län & SE4 & 0 & 0 & 0 \\ 
        \textbf{Tomelilla} & Skåne län & SE4 & 0 & 0 & 0 \\ 
        \textbf{Bromölla} & Skåne län & SE4 & 0 & 0 & 0 \\ 
        \textbf{Osby} & Skåne län & SE4 & 0 & 0 & 0 \\ 
        \textbf{Perstorp} & Skåne län & SE4 & 0 & 0 & 0 \\ 
        \textbf{Klippan} & Skåne län & SE4 & 0 & 0 & 0 \\ 
        \textbf{Åstorp} & Skåne län & SE4 & 0 & 0 & 0 \\ 
        \textbf{Båstad} & Skåne län & SE4 & 0 & 0 & 0 \\ 
        \textbf{Malmö} & Skåne län & SE4 & 0 & 0 & 0 \\ 
        \textbf{Lund} & Skåne län & SE4 & 0 & 0 & 0 \\ 
        \textbf{Landskrona} & Skåne län & SE4 & 0 & 0 & 0 \\ 
        \textbf{Helsingborg} & Skåne län & SE4 & 0 & 0 & 0 \\ 
        \textbf{Höganäs} & Skåne län & SE4 & 0 & 0 & 0 \\ 
        \textbf{Eslöv} & Skåne län & SE4 & 0 & 0 & 0 \\ 
        \textbf{Ystad} & Skåne län & SE4 & 0 & 0 & 0 \\ 
        \textbf{Trelleborg} & Skåne län & SE4 & 0 & 0 & 0 \\ 
        \textbf{Kristianstad} & Skåne län & SE4 & 0 & 0 & 0 \\ 
        \textbf{Simrishamn} & Skåne län & SE4 & 0 & 0 & 0 \\ 
        \textbf{Ängelholm} & Skåne län & SE4 & 0 & 0 & 0 \\ 
        \textbf{Hässleholm} & Skåne län & SE4 & 0 & 0 & 0 \\ 
        \textbf{Hylte} & Hallands län & SE4 & 0 & 0 & 0 \\ 
        \textbf{Halmstad} & Hallands län & SE4 & 0 & 0 & 0 \\ 
        \textbf{Laholm} & Hallands län & SE4 & 0 & 0 & 0 \\ 
        \textbf{Falkenberg} & Hallands län & SE4 & 0 & 0 & 0 \\ \hline 
        \label{tab:municipality_list}
\end{longtable}

\newpage
\subsection*{Additional figures and tables}

\begin{table}[htb!]
    \centering
    \small
    \caption{Variation within the SCGs}
    \label{tab:scg_variation}
    \begin{tabular}{lcccccccc} \hline
         & \multicolumn{3}{c}{\textbf{PV capacity}} & \textbf{Average} & \textbf{Radiation} & \textbf{Income} & \textbf{Housing} & \textbf{Unemploy-} \\
        & 2016 & 2018 & 2020 & \textbf{radiation} & \textbf{variation} & ~ & ~ & \textbf{ment} \\ \hline
        & ~ & ~ & ~ & ~ & ~ & ~ & ~ & ~ \\ 
        \textbf{SCG1} & ~ & ~ & ~ & ~ & ~ & ~ & ~ & ~ \\ 
        \textbf{Sandviken} & 0.004 & 0.007 & 0.048 &  1,208  &  28  &  206,000  & 53\% & 7.6\% \\ 
        \textbf{Vansbro} & 0.004 & 0.028 & 0.157 &  1,112  &  27  &  208,000  & 76\% & 8.2\% \\ 
        \textbf{Gagnef} & 0.010 & 0.040 & 0.085 &  1,160  &  29  &  208,000  & 90\% & 8.2\% \\ 
        \textbf{Borlänge} & 0.000 & 0.016 & 0.060 &  1,198  &  29  &  204,000  & 45\% & 8.2\% \\ 
        \textbf{} & ~ & ~ & ~ & ~ & ~ & ~ & ~ & ~ \\ 
        \textbf{SCG 2} & ~ & ~ & ~ & ~ & ~ & ~ & ~ & ~ \\ 
        \textbf{Gagnef} & 0.010 & 0.040 & 0.085 &  1,160  &  29  &  208,000  & 90\% & 8.2\% \\ 
        \textbf{Vansbro} & 0.004 & 0.028 & 0.157 &  1,112  &  27  &  208,000  & 76\% & 8.2\% \\ 
        \textbf{Norberg} & 0.016 & 0.048 & 0.093 &  1,214  &  28  &  186,000  & 53\% & 8.5\% \\ 
        \textbf{Borlänge} & 0.000 & 0.016 & 0.060 &  1,198  &  29  &  204,000  & 45\% & 8.2\% \\ 
        \textbf{Sandviken} & 0.004 & 0.007 & 0.048 &  1,208  &  28  &  206,000  & 53\% & 7.6\% \\ 
        \textbf{} & ~ & ~ & ~ & ~ & ~ & ~ & ~ & ~ \\ 
        \textbf{SCG 3} & ~ & ~ & ~ & ~ & ~ & ~ & ~ & ~ \\ 
        \textbf{Gagnef} & 0.010 & 0.040 & 0.085 &  1,160  &  29  &  208,000  & 90\% & 8.2\% \\ 
        \textbf{Vansbro} & 0.004 & 0.028 & 0.157 &  1,112  &  27  &  208,000  & 76\% & 8.2\% \\ 
        \textbf{Borlänge} & 0.000 & 0.016 & 0.060 &  1,198  &  29  &  204,000  & 45\% & 8.2\% \\ 
        \textbf{Falun} & 0.011 & 0.030 & 0.090 &  1,248  &  30  &  201,000  & 50\% & 7.3\% \\ 
        \textbf{Norberg} & 0.016 & 0.048 & 0.093 &  1,214  &  28  &  186,000  & 53\% & 8.5\% \\ 
        \textbf{Leksand} & 0.020 & 0.032 & 0.077 &  1,126  &  28  &  197,000  & 74\% & 7.3\% \\ \hline 
    \end{tabular}
\end{table}

\newpage
\subsection*{Difference-in-difference estimations}

The SCM estimations in this study are compared to a difference-in-difference (DiD) which has been applied by \cite{mauritzen2023great} to estimate the impact of electricity price divergence across BZ on electric vehicle uptake. 

The DiD rests on two main assumptions \citep{angrist2009mostly,zeldow2023}. First, according to the assumption of parallel trends divergence in the outcome variable should occur until the on-set of treatment, which in this study refers to tariff divergence across BZ in 2020. Second, nothing but treatment should affect outcome trends. Whether these assumptions hold can be evaluated with help of econometric methods or in case of the parallel trends assumptions by graphic data inspection \citep{billmeier2013assessing,zeldow2023}.

\begin{figure}[htb!]
    \centering
    \caption{Difference-in-difference}
    \includegraphics[scale=0.7]{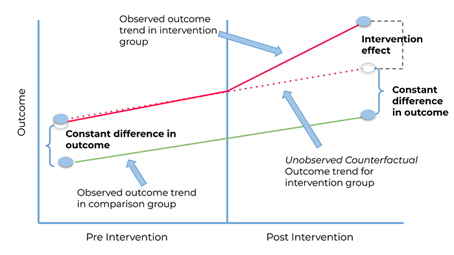}
    \label{fig:did_example} \\
    Source: Adapted from \cite{keisha2022}
\end{figure}

The main concepts of DiD is illustrated in figure \ref{fig:did_example}. During the pre-intervention period the outcomes of the treatment and control group parallel each other while the initial levels do not need to be similar. Only in the post treatment period, outcomes of the treatment and the control group are allowed to diverge. The actual treatment effect is then calculated as the divergence between the total outcome levels and initial variations from the pre-treatment period.

In the present study, the Northern and Southern BZ border regions are considered as the treated and control unit, respectively. The equation for the DiD estimation is presented in equation \ref{eq:did} \citep[p.187-189]{angrist2009mostly}. The outcome variable \textit{Y} refers to the installed PV potential per capita in a municipality \textit{i} during a given year \textit{t}. This outcome is determined by the value of the intercept \(\alpha\) and two dummy variables \textit{North} and \textit{Post}. \textit{North} is a dummy variable taking the value one for municipalities located in the low-price BZs SE1 or SE2, hence \(\beta\) estimates the effect of being located in the low-price zone on PV uptake. \textit{Post} indicates whether observations are collected during the post-treatment period. Consequentially, \(\gamma\) is included if observations are reported after the start of the treatment therefore in 2020 or later. The interaction term \(\delta\) measures if an observation is collected from a treated municipality during the post treatment period. \(\epsilon\) is an error term that captures both individual and time fixed effects. 

\begin{equation}
    \mathrm{Y_{it}=\alpha + \beta North_i + \gamma Post_t + \delta (North_i*Post_t)} + \epsilon_{it}
    \label{eq:did}
\end{equation}

Despite the possibilities to test underlying assumptions, \cite{billmeier2013assessing} finds that the parallel trends assumption is violated in several empirical studies. Moreover, DiD has the caveat that it lacks formal rules to select suitable control groups. Consequentially, the researchers choice of the treated and control units can bias the estimation results.

\subsection*{Estimation results}

\begin{figure}[htb!]
    \centering
    \caption{DiD - Border 1, 2016-2022 (kW)}
    \includegraphics[scale=0.55]{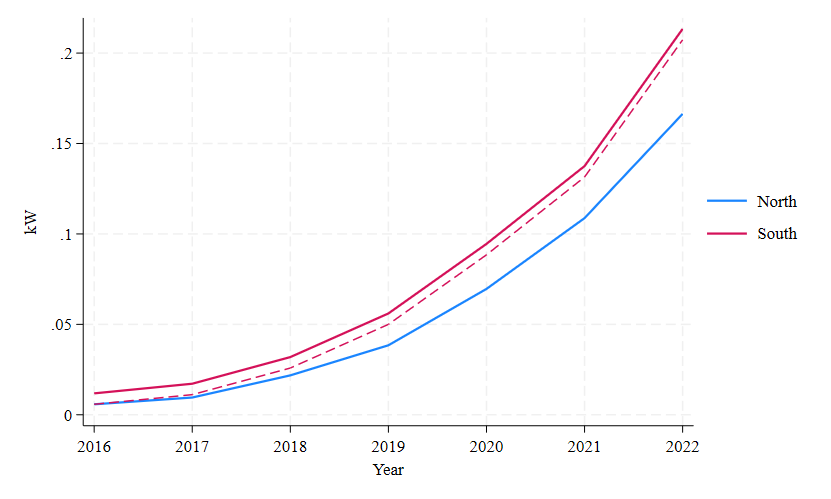}
    \label{fig:did_b1}
\end{figure}

\begin{figure}[htb!]
    \centering
    \caption{DiD - Border 3, 2016-2022 (kW)}
    \includegraphics[scale=0.55]{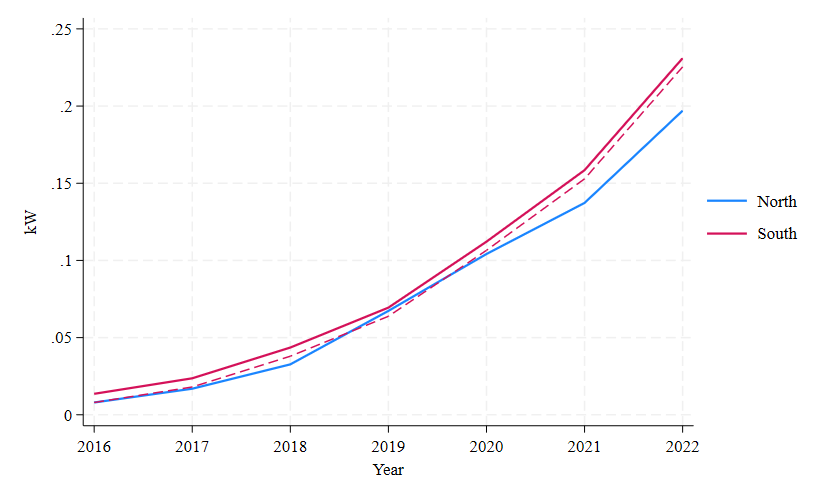}
    \label{fig:did_b3}
\end{figure}

The DiD regression results are presented in table \ref{tab:did_results} and in graphic form in figure \ref{fig:did_b1}, \ref{fig:border_2}, and \ref{fig:did_b3}. For each border region two separate estimation are displayed in- and excluding the control variables disposable income, housing structure and the unemployment rate as control variables. Radiation is not considered here as its value is within municipalities across time.

\begin{table}[htb!]
    \centering
    \caption{DiD - Regression results}
    \label{tab:did_results}
\begin{tabular}{lcccccc} 
\hline
            & \multicolumn{2}{c}{Border 1}  & \multicolumn{2}{c}{Border 2}  & \multicolumn{2}{c}{Border 3}  \\
            &      (1)      &      (2)      &      (3)      &     (4)       &       (5)     &       (6)     \\ \hline
            &               &               &               &               &               &               \\
ATET    &      -0.023***&      -0.009   &      -0.007   &       0.003   &      -0.015***&      -0.003** \\
            &      (0.00)   &         (.)   &         (.)   &         (.)   &      (0.00)   &      (0.01)   \\
Income      &               &      -0.000   &               &      -0.000   &               &      -0.000** \\
            &               &         (.)   &               &         (.)   &               &      (0.00)   \\
Housing      &               &      -0.167   &               &      -1.407   &               &      -0.562*  \\
            &               &         (.)   &               &         (.)   &               &      (0.04)   \\
Unemployment &               &      -0.005   &               &      -0.005   &               &       0.002*  \\
            &               &         (.)   &               &         (.)   &               &      (0.01)   \\
Constant       &       0.009   &       1.490   &       0.011   &       1.731   &       0.011*  &       1.534** \\
            &      (0.22)   &         (.)   &      (0.17)   &         (.)   &      (0.03)   &      (0.01)   \\
Year FE & Yes & Yes & Yes & Yes & Yes & Yes \\
 \hline
\multicolumn{7}{c}{ Robust standard errors in parentheses} \\
\multicolumn{7}{c}{ *** p$<$0.01, ** p$<$0.05, * p$<$0.1} \\
\end{tabular}
\end{table}

The size of the treatment effect estimate \textit{ATET} indicates how the sudden divergence in electricity prices affects installed solar PV capacity per capita in the Northern compared to the Southern BZ border in absolute terms. When control variables are excluded, the treatment effect is significant at the 1\% level for border 1 and 3. The amplitude of the treatment effect is largest for border 1. Being located in the low-tariff BZ reduced installed PV capacity by 0.0023 kW.

Estimation results including control variables are only significant for border 3. Nonetheless, the estimated treatment effect is with -0.003 very small and lacks economic significance. Moreover, the control variables have unexpected pre-signs. While disposable income, and the share of small houses are positively associated with PV uptake, and the relationship is negative for unemployment the opposite is the case here. 

Wile the vertical red line marks the beginning of the treatment period, the light green line is added to visualize in how far the outcome trend of the treatment and control unit are parallel to each other at the start of the treatment period.

For border 1, the Northern and Southern BZ border follow a similar trend in installed PV capacity over the treatment period but the absolute gap increases 0.006 to 0.047 kW per capita. Indicating that the parallel trends assumption might be violated. Moreover, divergence in installed PV potential already emerges in 2018, two years before the divergence in electric tariff rates. Therefore electricity price divergence might not be the sole factor driving divergence in PV uptake between the Northern and Southern border region.

\begin{figure}[htb!]
    \centering
    \caption{DiD - Border 2, 2016-2022 (kW)}
    \includegraphics[scale=0.55]{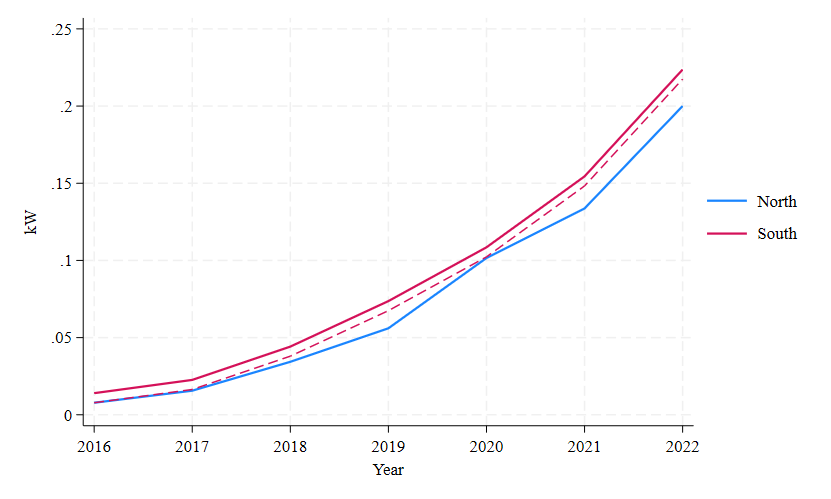}
    \label{fig:did_b2}
\end{figure}

A visual inspection of figures \ref{fig:did_b2} and \ref{fig:did_b3} indicates that the divergence in installed PV capacity is less pronounced when municipalities in greater geographical distance from the tariff border are considered. In this case there is especially no divergence in installed PV per capita levels before 2020.

To test whether the underlying assumptions are violated, a parallel trends and a placebo test are conducted. For the placebo test it is assumed that treatment already occurs in 2018, hence 2 year before the actual treatment

\begin{table}[htb!]
    \centering
    \caption{Parallel trends assumption}
    \label{tab:did_parallel}
\begin{tabular}{lccc} 
\hline
            & Border 1  & Border 2 & Border 3  \\ \hline
North       &       0.000   &       0.000   &       0.000   \\
            &       (.)     &         (.)   &      (.)      \\
South       &      -7.483   &     -7.480    &      1.293    \\
            &      (0.30)   &      (0.36)   &      (0.89)   \\
Year        &       0.011***&       0.016***&       0.019***\\
            &      (0.00)   &      (0.00)   &      (0.00)   \\
North x Year&       0.000   &      0.000    &     0.000     \\
            &       (.)     &         (.)   &      (.)      \\
South x Year&       0.004   &      0.004    &     -0.001    \\
            &      (0.30)   &      (0.36)   &      (0.90)   \\
Constant    &     -22.203***&     -32.942***&     -39.067***\\
            &      (0.00)   &      (0.00)   &      (0.00)   \\
 \hline
\multicolumn{4}{c}{ Robust standard errors in parentheses} \\
\multicolumn{4}{c}{ *** p$<$0.01, ** p$<$0.05, * p$<$0.1} \\
\end{tabular}
\end{table}

According to the results in table \ref{tab:did_parallel}, the assumption of parallel pre-treatment trends holds for all border definition. This is indicated by the insignificance of the interaction term between the time variable \textit{Year} and membership in the control group \textit{South}.

\begin{table}[htb!]
    \centering
    \caption{Placebo test}
    \label{tab:did_placebo}
\begin{tabular}{lcccccc} 
\hline
            & \multicolumn{2}{c}{Border 1}  & \multicolumn{2}{c}{Border 2}  & \multicolumn{2}{c}{Border 3}  \\
            &      (1)      &      (2)      &      (3)      &     (4)       &       (5)     &       (6)     \\ \hline
            &               &               &               &               &               &               \\
ATET    &      -0.019***&      -0.007   &      -0.009***&      -0.002   &      -0.009***&       0.019***\\
            &     (0.000)   &         (.)   &     (0.000)   &         (.)   &     (0.000)   &     (0.000)   \\
Income      &               &      -0.000   &               &      -0.000   &               &      -0.000***\\
            &               &         (.)   &               &         (.)   &               &     (0.001)   \\
Housing     &               &       1.417   &               &      -1.320   &               &       1.301** \\
            &               &         (.)   &               &         (.)   &               &     (0.006)   \\
Unemployment &               &      -0.005   &               &      -0.006   &               &       0.012***\\
            &               &         (.)   &               &         (.)   &               &     (0.001)   \\
Constant        &       0.009*  &       0.880   &       0.011*  &       1.420   &       0.011*  &       2.258** \\
            &     (0.042)   &         (.)   &     (0.017)   &         (.)   &     (0.024)   &     (0.001)   \\
Year FE & Yes & Yes & Yes & Yes & Yes & Yes \\
 \hline
\multicolumn{7}{c}{ Robust standard errors in parentheses} \\
\multicolumn{7}{c}{ *** p$<$0.01, ** p$<$0.05, * p$<$0.1} \\
\end{tabular}
\end{table}

Unlike the parallel pre-treatment trends, the placebo test results in table \ref{tab:did_placebo}, displays significant estimates for both specifications of border 3 and for border 1 and 2 are included. This suggests, that something else than tariff-divergence in 2020 is driving difference in PV uptake across BZ.

\end{document}